\documentstyle[12pt,epsf,epsfig]{article}

\newcommand{\beq}{\begin{equation}}
\newcommand{\eeq}{\end{equation}}
\newcommand{\beqa}{\begin{eqnarray}}
\newcommand{\eeqa}{\end{eqnarray}}
\newcommand{\no}{\nonumber}
\newcommand{\q}{\quad}
\newcommand{\qq}{\qquad}
\newcommand{\mnod}{\stackrel{\circ}{M}}
\newcommand{\Fnod}{\stackrel{\circ}{F}}
\newcommand{\tr}{\mbox{tr}}

\begin{document}

\hfill 

\hfill 

\bigskip\bigskip

\begin{center}

{{\Large\bf  Baryon axial currents }}

\end{center}

\vspace{.4in}

\begin{center}
{\large B. Borasoy\footnote{email: borasoy@het.phast.umass.edu }}

\bigskip

\bigskip

Department of Physics and Astronomy\\
University of Massachusetts\\
Amherst, MA 01003, USA\\

\vspace{.2in}

\end{center}

\vspace{.7in}

\thispagestyle{empty} 

\begin{abstract}
The baryon axial currents are calculated at one--loop order
in heavy baryon chiral perturbation theory by employing both a cutoff
and dimensional regularization. Data from the semileptonic baryon decays
is used to perform a least--squares fit to the two
axial couplings $D$ and $F$ of the effective Lagrangian. Predictions for
the momentum dependent axial form factors are made and
the two different regularization schemes are compared.
We also include the spin--3/2 decuplet in the effective theory.

\end{abstract}

\vspace*{\fill}

\section{Introduction}
In a recent work \cite{DHB} a cutoff regularization was introduced
in the framework of $SU(3)$ heavy baryon chiral perturbation theory.
Therein, it was demonstrated that short
distance effects, arising from propagation of Goldstone bosons over
distances smaller than a typical hadronic size, are model-dependent
and lead to a lack of convergence in the SU(3) chiral expansion when
regularized dimensionally.
The use of a cutoff removes such effects in a chirally consistent way
and improves the convergence problems which arise from large loop effects.
A simple dipole regulator was employed in that work, however,
the specific form of the cutoff is irrelevant 
for the cases discussed there -- a consistent chiral expansion can be carried
out.

We will analyze the
$SU(3)$ breaking corrections to the baryon axial currents
both by employing dimensional regularization and a cutoff.
First, this calculation serves as a comparison between 
both regularization schemes.
In addition, this generalizes other investigations  carried
out in dimensional regularization \cite{BSW,JM}. 
We will present the results for all three momentum--dependent axial 
form factors.
So far, only the
leading nonanalytic corrections to one of the
three axial form factors at zero momentum transfer were calculated
in the limit of vanishing light quark masses $m_u = m_d =0$.
The chiral logarithmic corrections to the axial currents
were as big as the lowest order values , and the authors came to the
conclusion that the axial couplings $D$ and $F$ of the lowest order
effective Lagrangian cannot be reliably extracted from
hyperon semileptonic decays \cite{JM}.

Another complication arises from the closeness of the spin--3/2 decuplet
resonances which are separated only by 231 MeV in average from the octet
baryons which is considerably smaller than the kaon and the $\eta$ mass.
These resonances are, therefore, expected to play an important part
at low energies. It has been suggested \cite{JM1} to include the
decuplet explicitely. There, it is shown that the spin--3/2 decuplet
partially cancels the large spin--1/2 octet contribution.
The average octet--decuplet mass splitting was chosen to be $\Delta=0$ MeV.
One has to reinvestigate this topic in the cutoff scheme and
also the impact of the parameter  $\Delta$ to the decuplet contributions.

The present work is organized as follows. 
In the next Section the definitions for the axial form factors
in the framework of heavy baryon chiral perturbation theory are presented
and related to the conventional relativistic formulation.
Results for the three axial form factors are shown.
Sec.~3 deals with the inclusion of the decuplet. A least--squares fit
for $D$ and $F$ is performed in Sec.~4. The chiral expansions 
of the hyperon semileptonic decays are presented. We conclude with a summary
in Sec.~5. The computation of the integrals is relegated to the Appendix.

\section{Axial form factors}
The hadronic axial current for the decay
$B_i \rightarrow B_j l \bar{\nu}_l$
can be written in the form
\beq
< B_j | A_{\mu} | B_i > =   \bar{u} (p_j) \bigg(
g_1 (q^2) \gamma_{\mu} \gamma_5 - \frac{ i \, g_2 (q^2) }{ M_i + M_j }
\sigma_{\mu \nu} q^{\nu} \gamma_5 + \frac{ g_3 (q^2) }{ M_i + M_j } q_{\mu} 
\gamma_5    \bigg) u(p_i)  \qq ,
\eeq
where $q = p_i - p_j $ is the momentum transfer. The form factor $g_3$
is usually neglected since after 
contraction with the leptonic current it is 
multiplied by a small lepton mass and therefore difficult to observe.
In this work, we calculate the three form factors in the framework 
of heavy baryon chiral perturbation theory both using dimensional
regularization and in a momentum dependent cutoff scheme.

The pseudoscalar Goldstone fields ($\phi = \pi, K, \eta$) are collected in
the  $3 \times 3$ unimodular, unitary matrix $U(x)$, 
\begin{equation}
 U(\phi) = u^2 (\phi) = \exp \lbrace 2 i \phi / \Fnod \rbrace
\end{equation}
with $\Fnod$ being the pseudoscalar decay constant (in the chiral limit), and
\begin{eqnarray}
 \phi =  \frac{1}{\sqrt{2}}  \left(
\matrix { {1\over \sqrt 2} \pi^0 + {1 \over \sqrt 6} \eta
&\pi^+ &K^+ \nonumber \\
\pi^-
        & -{1\over \sqrt 2} \pi^0 + {1 \over \sqrt 6} \eta & K^0
        \nonumber \\
K^-
        &  \bar{K^0}&- {2 \over \sqrt 6} \eta  \nonumber \\} 
\!\!\!\!\!\!\!\!\!\!\!\!\!\!\! \right) \, \, \, \, \, . 
\end{eqnarray}
Under SU(3)$_L \times$SU(3)$_R$, $U(x)$ transforms as $U \to U' =
LUR^\dagger$, with $L,R \in$ SU(3)$_{L,R}$.
One forms an object of axial--vector type with one derivative
\beqa
u_{\mu} & = & i u^{\dagger} \nabla_{\mu} U u^{\dagger} \no \\
\nabla_{\mu} U & = & \partial_{\mu} U - i a_{\mu} U - i U a_{\mu}
\eeqa
with $\nabla_{\mu}$ being the covariant derivative of $U$ 
and $ a_{\mu} $ generates the Green functions of the axial current. 
The matrix $B$ denotes the baryon octet, 
\begin{eqnarray}
B  =  \left(
\matrix  { {1\over \sqrt 2} \Sigma^0 + {1 \over \sqrt 6} \Lambda
&\Sigma^+ &  p \nonumber \\
\Sigma^-
    & -{1\over \sqrt 2} \Sigma^0 + {1 \over \sqrt 6} \Lambda & n
    \nonumber \\
\Xi^-
        &       \Xi^0 &- {2 \over \sqrt 6} \Lambda \nonumber \\} 
\!\!\!\!\!\!\!\!\!\!\!\!\!\!\!\!\! \right)  \, \, \, .
\end{eqnarray}
The matrices $u_{\mu}$ and $B$ transform under $SU(3)_L \times SU(3)_R$ 
as any matter field, ${\it e.g.}$,
\begin{equation} 
B \to B' = K \, B \,  K^\dagger
 \, \, \, ,
\end{equation}
with $K(U,L,R)$ the compensator field representing an element of the
conserved subgroup SU(3)$_V$.
In the heavy baryon formulation the baryons are described by a 
four--velocity $v_{\mu}$ and relativistic corrections appear as
$1/ \mnod$ corrections where $\mnod$ is the average octet baryon mass
in the chiral limit. A consistent chiral counting scheme emerges,
${\it i.e.}$ a one--to--one correspondence between the Goldstone boson loops
and the expansion in small momenta and quark masses.
To this end, one constructs eigenstates of the velocity projection
operator $P_v =  ( 1 + v\! \! /)/2$
\beq
B_v (x) = e^{i \mnod \, v \cdot x} \: P_v \, B (x) \qq.
\eeq
The Dirac algebra simplifies considerably. It allows to express any Dirac 
bilinear $\bar{B_v} \Gamma_{\mu} B_v \, (\Gamma_{\mu} = 1, \gamma_{\mu}, 
\gamma_5, \ldots ) $ in terms of the velocity $v_{\mu}$ and the spin operator
$ 2 S_{\mu} = i \gamma_5 \sigma_{\mu \nu} v^{\nu} $. The latter obeys the
relations (in $d$ space--time dimensions)
\beq
S \cdot v = 0 \q , \q S^2 = \frac{1-d}{4} \q , \q
\{ S_{\mu}, S_{\nu} \} = \frac{1}{2} ( v_{\mu} v_{\nu} - g_{\mu \nu} ) \q ,\q
[ S_{\mu}, S_{\nu} ] = i \epsilon_{\mu \nu \alpha \beta } v^{\alpha} S^\beta
\eeq
Using the convention $\epsilon^{0123} = 1$, one can rewrite the Dirac
bilinears as :
\beqa
\bar{B_v} \gamma_\mu B_v  & = &  v_\mu \bar{B_v}  B_v  \q , \q
\bar{B_v} \gamma_5 B_v   =   0 \q , \q
\bar{B_v} \gamma_\mu \gamma_5 B_v   =   2 \bar{B_v} S_{\mu} B_v  \q ,  \no \\
\bar{B_v} \sigma_{\mu \nu} B_v  & = &  2 \epsilon_{\mu \nu \alpha \beta } 
                v^{\alpha} \bar{B_v} S^\beta   B_v  \q , \q
\bar{B_v} \gamma_5 \sigma_{\mu \nu} B_v   =   2 i \, (
   v^{\mu} \bar{B_v} S^\nu   B_v  -  v^{\nu} \bar{B_v} S^\mu   B_v \, )  \q .
\eeqa
In the following, we will drop the index $v$.
In the heavy baryon formulation the hadronic axial current can be 
decomposed into
\beq
< B_j | A_{\mu} | B_i > =   \bar{H} (q_j) \bigg(
G_1 (q^2) \: S_{\mu} + \frac{ G_2 (q^2) }{ M_i + M_j } \: v_\mu \, S \cdot q 
+ \frac{ G_3 (q^2) }{ (M_i + M_j)^2 } \: q_{\mu} \, S \cdot q \, \bigg) 
    \:H(q_i)  \qq ,
\eeq
with $2H = (1 + v \! \! /)\; u$ being the large--component field and
$\: q_k = p_k - \mnod v \:$ the baryon off--shell momenta in the heavy baryon
formulation. We prefer to work in the rest frame of the heavy baryon
and with $ v_\mu = (1,0,0,0) $. The momenta of the baryons are then given
by
\beq
p_i = \mnod v + q_i = M_i v \: , \qq  p_j = \mnod v + q_j = M_i v - q \: .
\eeq
In this frame the $G_i$ are related to the $g_i$ via
\beqa \label{rel}
G_1(q^2)  & = &  2 \, g_1(q^2) + \frac{ g_2 (q^2) }{ M_i + M_j } \, \bigg(
        - v \cdot q + \frac{ (v \cdot q)^2 - q^2}{M_j + E_j} \bigg) \no \\ 
G_2(q^2)  & = &  2\, \frac{M_j + M_i}{M_j + E_j} \, g_1(q^2) 
           +2 \,g_2(q^2)\, \bigg(
          1 - \frac{ v \cdot q}{M_j + E_j} \bigg) \no \\ 
G_3(q^2)  & = & 2 \, \frac{M_j + M_i}{M_j + E_j} \, \bigg ( 
                   g_2(q^2) - g_3 (q^2) \bigg) \qq ,
\eeqa
with 
\beq \label{ener}
E_j = M_i - v \cdot q = M_i - \frac{1}{2 M_i} \Big( q^2 + M_i^2 - M_j^2 \Big)
\eeq
being the energy of the outgoing baryon.
There are contributions from $g_2$ to $G_1$  which
have been neglected in previous works \cite{JM,JM1}.

The Lagrangian can be decomposed into a purely mesonic part 
$ {\cal L}_{\phi }$ and a piece ${\cal L}_{\phi \, B}$ in which 
the pseudoscalar Goldstone bosons are coupled to the baryon fields
\beq
 {\cal L} =  {\cal L}_{\phi \, B} +  {\cal L}_{\phi} \qq .
\eeq  
For the mesonic part one has \cite{GL}
\beq
{\cal L} = \frac{\Fnod^2}{4} \, \tr \Big[ u_{\mu} u^{\mu} \Big] +
           \frac{\Fnod^2}{4} \, \tr \Big[ \chi_+ \Big]  \qq ,
\eeq
with $\chi_+ = 2 B_0 (u^\dagger {\cal M} u^\dagger + u {\cal M} u )$
being proportional to the quark mass matrix ${\cal M} = \mbox{diag} (m_u,
m_d,$ $ m_s)$. Also, $B_0 = - \langle 0 | \bar{q} q | 0 \rangle / \Fnod^2$ is
the order parameter of the spontaneous symmetry violation,
and we assume $B_0 \gg \Fnod$. 
In the following we will work in the isospin limit
$m_u = m_d$.

The lowest order baryon meson Lagrangian 
$ {\cal L}^{(1)}_{\phi \, B}$ includes the 
two axial--vector couplings $D$ and $F$
\beq
{\cal L}^{(1)}_{\phi \, B} = \, i \, \tr \Big( \bar{B} 
       [ v \cdot D , B] \Big) + 
   D \, \tr \Big( \bar{B} S_{\mu} \{ u^{\mu}, B\} \Big) 
  + F \, \tr \Big( \bar{B} S_{\mu} [ u^{\mu}, B] \Big) \qq ,
\eeq
and the superscrpit denotes the chiral order.
The covariant derivative $D_{\mu}$ on the baryon fields includes
the external gauge field $a_{\mu}$ :
\beqa
[ D_{\mu}, B ]  & = &  \partial_\mu B + [ \Gamma_{\mu}, B] \no \\
\Gamma_\mu   & = &   \frac{1}{2} [ u^\dagger, \partial_\mu u ] -
  \frac{i}{2} ( u^\dagger a_\mu u  - u a_\mu u^\dagger )
\eeqa
with $ \Gamma_\mu $ the so--called chiral connection. 
If only this effective Lagrangian would be 
used to compute the chiral corrections in the effective theory, there would
be no contributions to the form factor $G_2$. From the second equation in
(\ref{rel}) this would lead to $g_1[pn](0) = -g_2(0)[pn] \simeq 1.26$
for the decay $ n \rightarrow p$.
This is in sharp contradiction with the value $g_2[pn](0)= 0$ assuming
$G$--parity. Therefore, we will also include  the terms from
the baryon meson Lagrangian at second chiral order that contribute 
at tree level to the form factors
\beqa
{\cal L}^{(2)}_{\phi \, B} &=&\,\frac{i D}{2 \mnod} \tr \Big([D^{\mu},\bar{B}] 
          S_{\mu} \{ v \cdot u , B \} \Big) +
     \,\frac{i F}{2 \mnod} \tr \Big([D^{\mu},\bar{B}] 
          S_{\mu} [ v \cdot u , B ] \Big) \no \\
 & - &  \,\frac{i D}{2 \mnod} \tr \Big(\bar{B}
          S_{\mu} \{ v \cdot u , [D^{\mu}, B ]\} \Big) -
\,\frac{i F}{2 \mnod} \tr \Big(\bar{B}
          S_{\mu}  [ v \cdot u , [D^{\mu}, B ]] \Big) \qq.
\eeqa
These are relativistic corrections to the lowest order Lagrangian
and the only terms contributing to the axial currents at this order.
In the initial--baryon restframe one has the relation $S \cdot q_1 = 0$ and
the last two terms do not contribute.

Corrections from the Goldstone boson loops contribute at third chiral order
together with counterterms from the Lagrangian ${\cal L}_{\phi B}^{(3)}$.
The Lagrangian ${\cal L}_{\phi B}^{(3)}$ contains numerous counterterms
which contribute to the axial form factors \cite{BH}. Performing the 
calculation with the complete Lagrangian up--to--and--including
one--loop order one has, of course, no predicitve power.
The theoretical predictions contain considerably more low--energy
constants than there are experimental results. One can resort to
model dependent estimations of these constants, {\it e.g.} via 
resonance saturation. This method works very accurately in the meson sector
\cite{Ec}, but in the baryon case it gives only a rough estimate of the 
LECs and there is still a sizeable uncertainty in these parameters
\cite{BM}. Since such LECs renormalize the part of the loops analytic
in the quark masses, no reliable estimate can be given for the
analytic pieces at this order. We will therefore restrict ourselves
to the nonanalytic pieces of the one--loop integrals and completely neglect
the local counterterms at this order.

The leading nonanalytic corrections from the
Goldstone boson loops are usually evaluated by using dimensional
regularization \cite{BSW, JM, JM1}. The nonanalytic part of a typical
integral in this analysis has for the case
of zero momentum transfer and $d$ dimensions the form
\beq  \label{int}
\int  \frac{ d^d l }{(2 \pi)^d} \, \, \frac{ i^3 \, ( S \cdot l )^2}{
       [l^2 - m_{\phi}^2 + i \epsilon] \, [ v \cdot l + i \epsilon ]^2 } \:
       =  \: - \frac{3}{64 \pi^2} \, m_{\phi}^2 \ln 
                 \frac{ m_{\phi}^2}{\lambda^2} \q , 
\eeq
where $m_{\phi}$ is the meson mass and $\lambda$ the scale introduced
in dimensional regularization.
The integral grows  with increasing meson mass.
We expect the long distance portion of the integral to be larger
for small meson masses since for small momenta the meson propagator
can be approximated by $1/m_{\phi}^2$.
This indicates that in the dimensionally regularized integral
there are significant contributions from short distance physics
which cannot be described appropriately by chiral symmetry.
Therefore, one has to employ other regularization schemes that emphasize
long distance effects of the integrals and reduce short distance
contributions. In \cite{DHB} it was shown that a simple dipole
regulator fulfills these requirements.

For the evaluation of the Goldstone boson loops in the cutoff scheme
we will employ the dipole regulator
\beq
R = \bigg( \frac{ \Lambda^2}{\Lambda^2 - l^2} \bigg)^2  \q ,
\eeq
where $l$ is the loop momentum.
Inserting this regulator into the integral in Eq.~(\ref{int}) leads to
\beqa  \label{intlam}
I_{\Lambda} &=&
\int  \frac{ d^4 l }{(2 \pi)^4} \, \, \frac{ i^3 \, ( S \cdot l )^2}{
       [l^2 - m_{\phi}^2 + i \epsilon] \, [ v \cdot l + i \epsilon ]^2}
       \bigg( \frac{ \Lambda^2}{\Lambda^2 - l^2} \bigg)^2  \no \\
  &=&  \:  - \frac{3}{64 \pi^2} {\Lambda^4\over (\Lambda^2-m_\phi^2)^2}\bigg(
       \Lambda^2-m_\phi^2 + m_\phi^2 \ln{m_\phi^2\over\Lambda^2} \bigg) \qq .
\eeqa
The introduction of the additional scale $\Lambda$ spoilt the
one--to--one correspondence between the meson loops and the expansion
in the quark masses
and the integral depends strongly on the value of the cutoff $\Lambda$.
However, this does not mean that to the order
we are working the resulting physics will depend
on $\Lambda$, since one is able to absorb the effects of $\Lambda$ into
a renormalization of the couplings $D$ and $F$ as we will show later.

We can now proceed in writing down the results both for the case of
dimensional regularization and the regularization with the cutoff.
Note, that in \cite{BSW, JM} only the leading nonanalytic pieces for the
form factor $g_1(0)$ at zero momentum transfer have been calculated.

At tree level the contributing diagrams are shown in Fig.~1. The
diagrams 1.a and 1.b contribute to $G_{1,2}$ and $G_3$, respectively.
The coefficients for $G_1$ read
\begin{eqnarray}
G_1^{1+i2}[pn]   &=& \alpha_{pn} =  2( D+F)      \no\\
G_1^{1+i2}[\Lambda\Sigma^-]   &=& \alpha_{\Lambda\Sigma^-} =
                       {4\over \sqrt{6}} D   \no \\
G_1^{1+i2}[\Xi^0\Xi^-]   &=&  \alpha_{\Xi^0\Xi^-} = 2(D-F)   \no \\
G_1^{4+i5}[p\Lambda]   &=&  \alpha_{p\Lambda} = 
                       -{2\over \sqrt{6}} (D+3F)  \no \\
G_1^{4+i5}[\Lambda\Xi^-]  &=&  \alpha_{\Lambda\Xi^-} =
                            -{2\over \sqrt{6}}  (D-3F)  \no \\
G_1^{4+i5}[n\Sigma^-]    &=&   \alpha_{n\Sigma^-} =  2 (D-F)   \no \\
G_1^{4+i5}[\Sigma^0\Xi^-]   &=&  \alpha_{\Sigma^0\Xi^-} = \sqrt{2}  (D+F)
   ={1\over \sqrt{2}}G_1^{4+i5}[\Sigma^+\Xi^0] \qq .
\end{eqnarray}
For $G_2$ one obtains contributions only from
$ {\cal L}^{(2)}_{\phi \, B} $ with
\beq
G_2 [ij]   =  G_1 [ij] \qq ,
\eeq
where we replaced the appearing prefactor $(M_i + M_j)/\mnod$ in
each of the decays by 2 which is consistent to the order we are working.
At tree level we obtain from Eq.~(\ref{rel}), {\it e.g.}, 
$g_2[pn](0)=0$ which is consistent with the assumption of $G$--parity.
The relativistic corrections at next--to--leading order in the 
baryon meson Lagrangian play an important role for the form factors $g_2$.
For $G_3$ the results read
\beq
G_3 [ij] (q^2)  = \frac{ (M_i + M_j)^2 }{ m_\phi^2 - q^2} G_1 [ij]
\eeq
where $m_\phi = m_\pi$ for the decays $[ij] = [pn], [\Lambda\Sigma^-],
[\Xi^0\Xi^-]$ and $m_\phi = m_K$ otherwise. Here, we replaced the 
lowest order expression for the meson masses in the propagator
by their physical value. 
The difference shows up at second chiral order for the form factors, 
{\it i.e} the same order 
as the loop contributions, and can be incorporated into these, as we will
show.
We also neglected the contributions
from $ {\cal L}^{(2)}_{\phi \, B} $ to $G_3$ since they are proportional to
$v \cdot q$ and amount to analytical corrections at loop
order after employing Eq.~(\ref{ener}). 
It is these forms for $G_1$ which are used in SU(3) fits 
to hyperon beta decay.

The loop contributions for $G_1(q^2)$ are depicted in Fig.~2.
They have the form
\beq
\delta G_1 [ij] (q^2)  =  \frac {1}{\Lambda_\chi^2} \: 
       \sum_{\phi = \pi,K,\eta} \bigg(
      \Big[ \beta_{ij}^{\,\phi} - \frac{1}{2} \alpha_{ij} 
      (\lambda_i^{\,\phi}+\lambda_j^{\,\phi})
      \, \Big] \, I^{\,\phi}(0) + \gamma_{ij}^{\,\phi} \, 
      I^{\,\phi}(\frac{-q^2}{2M_i})
      \bigg) \, + \, \alpha_{ij} \delta_j (q^2)
\eeq
with $\Lambda_\chi = 4 \pi F_\pi$ and we have also included the wavefunction 
renormalization factors of the external baryons proportional to $\alpha_{ij} $.
Furthermore, one has to account for the contributions of the heavy
components of the external baryons to their $Z$-factors, see \cite{EM}.
In the rest frame of the heavy baryon they vanish for the
decaying baryon. For the light baryon with the mass $M_j$ we get a
term $\delta_j$ which is to lowest order proportional to $q^2 / (4 M_j^2)$. 
Also, $\Fnod$ has been replaced by $F_\pi$ which is consistent to the order
we are working, and $I^{\phi}$ is the integral appearing in the calculation.
The pertinent coefficients read

\begin{eqnarray}
\beta_{pn}^\pi &=&   -2 ( D + F) \q , \q
\beta_{pn}^K =  - (D+F)  \q , \q
 \beta_{pn}^\eta = 0 \no\\
\beta_{p\Lambda}^\pi &=& {\sqrt{6}\over 8} (D+3F) \q , \q
\beta_{p\Lambda}^K  =  { \sqrt{6} \over4} (D+3F)\q , \q
\beta_{p\Lambda}^\eta =  {\sqrt{6}\over 8} (D+3F) \no\\
\beta_{\Lambda\Sigma^-}^\pi &=& - {4\over \sqrt{6}} D \q , \q
\beta_{\Lambda\Sigma^-}^K  =  - {2\over \sqrt{6}} D \q , \q
\beta_{\Lambda\Sigma^-}^\eta = 0 \no\\
\beta_{\Xi^0\Xi^-} ^\pi  &=&  -2 (D-F)  \q , \q
\beta_{\Xi^0\Xi^-} ^K  =  - (D-F)  \q , \q
\beta_{\Xi^0\Xi^-} ^\eta  =  0   \no\\
\beta_{n\Sigma^-}^\pi  &=& -{3\over4}(D-F) \q , \q
\beta_{n\Sigma^-}^K  = -{3\over 2} (D-F) \q , \q
\beta_{n\Sigma^-}^\eta  =  -{3\over4}(D-F) \no\\
\beta_{\Lambda\Xi^-}^\pi  &=&  {\sqrt{6}\over 8}(D-3F) \q , \q 
\beta_{\Lambda\Xi^-}^K  = {\sqrt{6}\over 4}(D-3F) \q , \q 
\beta_{\Lambda\Xi^-}^\eta  = {\sqrt{6}\over 8}(D-3F) \no\\
\beta_{\Sigma^0\Xi^-}^\pi &=& -{3\sqrt{2} \over 8}(D+F) \q , \q 
\beta_{\Sigma^0\Xi^-}^K  =   -{3\sqrt{2} \over 4}(D+F) \q , \q
\beta_{\Sigma^0\Xi^-}^\eta  =  -{3\sqrt{2} \over 8}(D+F) \no\\
\beta_{\Sigma^+\Xi^0}^\phi &=& \sqrt{2} \beta_{\Sigma^0\Xi^-}^\phi \qq .
\eeqa
\beqa
\gamma_{pn}^\pi &=& {1\over 2}(D^3+F^3+3D^2F+3F^2D) \q , \q
\gamma_{pn}^K   = {2\over 3}D^3-{2\over 3}FD^2+2DF^2-2F^3 \q , \no \\
\gamma_{pn}^\eta &=&-{1\over 6}D^3+{5\over 6}FD^2 -{1\over
                      2}DF^2-{3\over 2}F^3 \q , \q
\gamma_{p\Lambda}^\pi  = {1\over \sqrt{6}}(-3 D^3+ 3 DF^2)\q , \q\no \\
\gamma_{p\Lambda}^K &=& {1\over \sqrt{6}}({5\over
                        3}D^3-5D^2F-3F^2D+9F^3) \q , \q
\gamma_{p\Lambda}^\eta  = {1\over \sqrt{6}}({1\over3}D^3-3DF^2) \q , \q\no\\
\gamma_{\Lambda\Sigma^-}^\pi &=& {4\over \sqrt{6}}
                              (-{1\over3}D^3+2DF^2) \q , \q
\gamma_{\Lambda\Sigma^-}^K  = {2\over\sqrt{6}}(D^3-DF^2)\q , \q\no\\
\gamma_{\Lambda\Sigma^-}^\eta  &=& {4\over 3 \sqrt{6}} D^3 \q , \q 
\gamma_{\Xi^0\Xi^-} ^\pi   =  {1\over 2}(D^3-F^3-3D^2F+3F^2D) \q , \q  \no\\
\gamma_{\Xi^0\Xi^-} ^K  &=&  {2\over 3}D^3+{2\over 3}FD^2+2DF^2+2F^3 \q , \q
\gamma_{\Xi^0\Xi^-} ^\eta  =  -{1\over 6}D^3-{5\over 6}FD^2 -{1\over
                      2}DF^2+{3\over 2}F^3 \q , \q  \no\\
\gamma_{n\Sigma^-}^\pi  &=& {1\over 3}D^3-{2\over 3}D^2F+DF^2+2F^3  \q , \q
\gamma_{n\Sigma^-}^K   = F^3+DF^2+{1\over
                       3}D^2F+{1\over 3}D^3  \q , \q \no\\
\gamma_{n\Sigma^-}^\eta   &=& DF^2-{4\over
                         3 }D^2F+{1\over 3}D^3  \q , \q
\gamma_{\Lambda\Xi^-}^\pi  =  {3\over \sqrt{6}}(-D^3+F^2D) \q , \q\no\\
\gamma_{\Lambda\Xi^-}^K  &=&  {1\over \sqrt{6}}({5\over
                          3}D^3+5D^2F-3DF^2-9F^3 )\q , \q 
\gamma_{\Lambda\Xi^-}^\eta  = {1\over
                       \sqrt{6}}({1\over 3}D^3-3DF^2) \q , \q\no\\
\gamma_{\Sigma^0\Xi^-}^\pi &=&   \sqrt{2} (-F^3+{1\over
                         3}FD^2+{1\over 2}F^2D+{1\over 6}D^3)  \q , \q \no \\
\gamma_{\Sigma^0\Xi^-}^K  &=&  \sqrt{2} ({1\over
           6}D^3-{1\over 6}FD^2+{1\over 2}F^2D-{1\over 2}F^3) \q , \q
\gamma_{\Sigma^0\Xi^-}^\eta  =  \sqrt{2} ({1\over 6}D^3+{2\over
                      3}D^2F+{1\over 2}DF^2) \q , \q \no \\
\gamma_{\Sigma^+\Xi^0}^\phi &=& \sqrt{2} \gamma_{\Sigma^0\Xi^-}^\phi \q .
\end{eqnarray}
The coefficients of the $Z$--factors are
\begin{eqnarray}
\lambda_N^\pi&=&{9\over4}(D+F)^2 \q ,\q 
\lambda_N^K={1\over2}(5D^2-6DF+9F^2) \q ,\q 
\lambda_N^\eta={1\over 4}(D-3F)^2  \q , \q \no\\
\lambda_\Sigma^\pi&=& D^2+6F^2  \q , \q
\lambda_\Sigma^K=3(D^2+F^2) \q ,\q
\lambda_\Sigma^\eta=D^2  \q ,\q \no\\
\lambda_\Lambda^\pi&=&3D^2 \q ,\quad
\lambda_\Lambda^K=D^2+9F^2 \q ,\quad
\lambda_\Lambda^\eta=D^2 \q , \nonumber\\
\lambda_\Xi^\pi&=&{9\over4}(D-F)^2 \q ,\quad
\lambda_\Xi^K={1\over2}(5D^2+6DF+9F^2) \q ,\quad
\lambda_\Xi^\eta={1\over 4}(D+3F)^2 \q . \no \\
\end{eqnarray}
The $\delta_{j}$ have the form
\beq
\delta_{j} = \frac{q^2}{8 M_j^2} \qq .
\eeq
In the integral, we set the off--shell momenta of the external baryons
$v \cdot q_i = 0$ and $v \cdot q_j = -q^2/(2 M_i^2)$ 
by neglecting the baryon mass differences which are of second chiral order.
This is consistent to the order we are working.
Then, the integral reads in the cutoff scheme for zero momentum
transfer squared $q^2 =0$
\beq \label{I1}
I^{\, \phi}(q^2=0) =
   \:   {\Lambda^4\over (\Lambda^2-m_{\phi}^2)^2}\bigg(\Lambda^2
      -m_{\phi}^2 + m_{\phi}^2 \ln {m_{\phi}^2\over\Lambda^2} \bigg) \qq .
\eeq
In dimensional regularization we obtain for the nonanalytic part of the
integral
\beq 
I_{dim}^{\, \phi}(q^2=0) =
  \:  m_{\phi}^2 \ln {m_{\phi}^2\over\lambda^2}  \qq .
\eeq
The more general cases for $q^2 =0$ can be found in App.~A.

As we mentioned above, the integral in Eq.(\ref{I1}) 
depends on $\Lambda$. But the physics
does not depend on the cutoff $\Lambda$. To this end, one expands
the result in Eq.(\ref{I1}) in terms of the meson mass $m_{\phi}$
\begin{equation}
I^{\, \phi} \stackrel{m_{\phi}^2 <<\Lambda^2}{\longrightarrow}
     \Lambda^2+m_{\phi}^2\ln {m_{\phi}^2\over \Lambda^2}+\ldots \qq .
\end{equation}
The second term delivers the nonanalytic contribution in dimensional
regularization. The contributions quadratic in $\Lambda$ can be absorbed
into renormalizations of the lowest order axial couplings $D$ and $F$ via
\begin{eqnarray}
D^r&=&D-{3\over 2}D(3D^2+5F^2+1){\Lambda^2\over 16\pi^2 F_\pi^2} \no \\
F^r&=&F-{1\over 6}F(25D^2+63F^2+9){\Lambda^2\over 16\pi^2 F_\pi^2} \qq .
\end{eqnarray}
Since such coefficients are determined empirically the analysis with
small meson masses becomes
identical to that of the dimensionally regularized case.
That the renormalization can occur involves a highly constrained set
of conditions and the fact that they are satisfied is a significant
verification of the chiral invariance of the cutoff procedure.

When employing any regularization scheme that introduces a dimensionful 
parameter, the usual power counting will be upset. This is manifest
in the results quoted above, in which the lower order chiral
parameters, $D$ and $F$, are shifted by the loop correction. However,
since these shifts are just the renormalization of phenomenological
parameters, they do not influence the physics. One can use the small
mass limit to set up the chiral expansion. In this limit the loops
will renormalize the chiral parameters and the power counting for
the order of the residual loop effects remains the same as in the
standard regularization, so that the same loop diagrams should be included
to a given order. Taking the meson masses to their physical values, which 
are in general not small compared to the cutoff, the short distance parts
of the loops will be discarded. In the calculation this amounts to a
partial resummation of higher order terms in the chiral expansion,
{\it i.e.} higher powers of $m_\phi/\Lambda$.

We prefer to remove the asymptotic mass--independent component of the
integral $I$ by setting
\beq
\tilde{I} = I - \Lambda^2
\eeq
and redefining the axial couplings $D$ and $F$. In the following we will work
with the renormalized values of $D$ and $F$ and neglect the
superscript $r$. 
The leading chiral corrections from the loops read then in the cutoff scheme
\beq
\delta G_1 [ij] (q^2)  =  \frac {1}{\Lambda_\chi^2} \: 
        \sum_{\phi = \pi,K,\eta} \bigg(
      \Big[ \beta_{ij}^{\,\phi} - \frac{1}{2} \alpha_{ij} 
      (\lambda_i^{\,\phi}+\lambda_j^{\,\phi})
      \, \Big] \, \tilde{I}^{\,\phi}(0) + \gamma_{ij}^{\,\phi} \, 
      \tilde{I}^{\,\phi}(\frac{-q^2}{2 M_i}) 
       \bigg) \, + \, \alpha_{ij} \delta_j (q^2) \q .
\eeq
In order to keep the formulae compact we use the same 
notation in the case of dimensional regularization with
\beq
\tilde{I}_{dim}^{\,\phi} = I_{dim}^{\,\phi} \q .
\eeq

The contributing loop diagrams to the form factor $G_3$ are shown in Fig.~3.
The graph 3.d leads to the meson $Z$--factor and the
renormalization of the meson mass in the propagator of the tree diagram 1.b.
Due to this diagram we used the physical value of the meson mass
at tree level.
Here, we neglect the analytic corrections to the meson mass from the 
counterterms at fourth chiral order of the mesonic Lagrangian. 
In our case, their contribution to the form factors turns out to be negligible.
One ends up with a similar form for the form factors $G_3$
\beqa
\delta G_3 [ij] (q^2)  &=&  \frac{(M_i + M_j)^2}{m_{\tilde{\phi}}^2  
        -q^2}\: \Bigg[
      \frac{1}{\Lambda_\chi^2} \:    \sum_{\phi = \pi,K,\eta} \bigg(
      \Big[ \frac{5}{3} \beta_{ij}^{\,\phi} - \frac{1}{2} \alpha_{ij} 
      (\lambda_i^{\,\phi}+\lambda_j^{\,\phi})
      \, \Big] \, \tilde{I}^{\,\phi}(0) + \gamma_{ij}^{\,\phi} \, 
      \tilde{I}^{\,\phi}(\frac{-q^2}{2 M_i}) \bigg) \no \\
    & & \qq \qq \q \, + \: \alpha_{ij}\:  (\: \delta_j (q^2) 
       + \zeta_{\tilde{\phi}} \: ) \: \Bigg] \q ,
\eeqa
with $\tilde{\phi} = \pi$ for the decays $[ij] = [pn], [\Lambda\Sigma^-],
[\Xi^0\Xi^-]$ and $\tilde{\phi} = K$ otherwise. The term 
$\zeta_{\tilde{\phi}}$ is due to the meson $Z$--factor and reads
\beqa
\zeta_{\,\pi} &=& \frac{1}{96 \pi F_\pi^2} \bigg( 4 \tilde{I}^{\,\pi}(0) +
                                   2 \tilde{I}^{\,K}(0) \bigg) \no \\
\zeta_{\,K} &=& \frac{1}{96 \pi F_\pi^2} 
          \bigg( \frac{3}{2} \tilde{I}^{\,\pi}(0) +
        3 \tilde{I}^{\,K}(0) + \frac{3}{2} \tilde{I}^{\,\eta}(0) \bigg) \q .
\eeqa

There are no loop contributions to $G_2$ at the order we are working. \\
Before presenting the numerical results we will include the decuplet
in the next section.

\section{Inclusion of the decuplet}
In general it is assumed that baryon resonance states are much heavier 
compared to the lowest--lying baryon octet. In this case they can be
integrated out and replaced by counterterms that do not include these
resonance states explicitely. However, while this might be a reasonable
procedure for heavier resonances like the Roper--octet, it is a
questionable assumption for the decuplet.
The low--lying decuplet is separated from the octet by only $\Delta = 231$
MeV in average which is much smaller than the $K$ or the $\eta$ mass. 
Furthermore, the $\Delta(1232)$ couples strongly to the $\pi N$ sector
and its contribution plays an important role in the channels wherein
this effect is possible.
In the meson sector, the first resonance is the vector meson $\rho$ with
a mass of 770 MeV which is considerably heavier than the
Goldstone bosons.
It was therefore argued in \cite{JM1} to include the spin--3/2 decuplet
as explicit degrees of freedom.
In the framework of conventional heavy baryon CHPT  it was shown that the
spin--3/2 decuplet partially cancels the large spin--1/2 octet contribution
for the form factor $g_1(0)$\cite{JM1}. 
This calculation was performed in the $SU(6)$ limit by
neglecting the octet--decuplet mass splitting $\Delta$ and also, $m_u =
m_d = 0$ was assumed.

In this section we will include the decuplet fields both in dimensional
regularization and in the cutoff scheme. We keep the average 
octet--decuplet mass splitting $\Delta$ at its physical value. 

The pertinent interaction Lagrangian between the spin--3/2 fields
-- denoted by the Rarita--Schwinger fields $T_{\mu}$ --, the baryon octet
and the Goldstone bosons reads 
\beq
{\cal L}_{\phi B T} = - i \, \, \bar{T}^{\mu} \, v \cdot D \, T_{\mu} \:
          + \Delta \,   \bar{T}^{\mu} \,   T_{\mu} \:           \:
          + \frac{C}{2} \, \Big( \,\bar{T}^{\mu} \, u_{\mu} B \, 
                     + \, \bar{B} \, u_{\mu}\, T^{\mu}\: \Big) 
          + H \, \bar{T}^{\mu} \, S_\nu  u^\nu \, T_{\mu}    
\eeq
where we have suppressed the flavor $SU(3)$ indices. In the heavy mass
formulation the fields $T_{\mu}$ satisfy the condition $ v \cdot T = 0$. 
The coupling constant $C = 1.2 ... 1.8$ can be determined from the
strong decays $T \rightarrow B \pi$. 
The parameter $H$ is not determined experimentally. Thus a fit to 
semileptonic hyperon decays involves one additional parameter
when intermediate decuplet states are included.
After integrating out the heavy degrees of freedom from the relativistic
Lagrangian there is still a remaining mass dependence which is proportional
to the average octet--decuplet splitting $\Delta$ and does not vanish
in the chiral limit.
In the Feynman rules the mass splitting $\Delta$ is contained in the decuplet
propagator
\beq
\frac{i}{ v \cdot l \, \, - \Delta + \, \, i \epsilon}      
\, \, \bigg( v_{\mu} v_{\nu} - g_{\mu \nu} 
-4 \frac{d-3}{d-1}S_{\mu}S_{\nu} \bigg)
\eeq
in $d$ dimensions.
The appearance of the mass scale $\Delta$ destroys in the case of
dimensional regularization the one--to--one correspondence between
meson loops and the expansion in small momenta and quark masses.
No further complications arise in our case since the strict chiral
counting scheme has already been spoilt by introducing the scale $\Lambda$.

The decuplet contributions to $G_1$ are presented in Fig.~4 and
have the following form
\beqa
\delta G_1 [ij] (q^2) &=& - \frac{C^2}{\Lambda_\chi^2}\:
      \sum_{\phi=\pi,K,\eta} \: \bigg( \frac{4}{3} \kappa_{ij}^{\, \phi} 
         J^{\, \phi}( -\Delta,-\frac{q^2}{2 M_i}) +\frac{4}{3} 
       \rho_{ij}^{\, \phi} J^{\, \phi}(0,-\frac{q^2}{2 M_i}-\Delta) \no \\
& & \qq \q
  -\frac{10}{9}  H \:\sigma_{ij}^{\, \phi} 
     J^{\, \phi}(- \Delta,-\frac{q^2}{2 M_i}-\Delta)
   - \alpha_{ij} (\epsilon_i^{\, \phi}  + \epsilon_j^{\, \phi} )
      J^{\, \phi}(- \Delta,-\Delta)     \bigg) \; . 
\eeqa
The integrals $J^{\, \phi}(x,y)$ can be found in App.~A and we neglected again
the baryon mass differences in the integrals by setting $v \cdot q_i =0$
and $v \cdot q_j = - q^2/(2 M_i)$. One also has to include the contributions
from intermediate states to the baryon $Z$--factors. The last term with
the coefficients 
$\epsilon_{i}^{\, \phi}$ takes this into account.
The coefficients read
\beqa
\kappa_{pn}^{\pi}   & = &  \frac{4}{3} ( D+F) \q , \q 
\kappa_{pn}^K    =   \frac{1}{2} D + \frac{1}{6} F \q , \q 
\kappa_{pn}^{\eta}    =   0  \no \\
\kappa_{\Lambda \Sigma^-}^{\pi}   & = &  - \frac{2}{3 \sqrt{6}}  D \q , \q 
\kappa_{\Lambda \Sigma^-}^K  = \frac{1}{\sqrt{6}} (\frac{5}{3} D + 3F) \q , \q 
\kappa_{\Lambda \Sigma^-}^{\eta}    =  \frac{1}{\sqrt{6}} D  \no \\
\kappa_{\Xi^0 \Xi^-}^{\pi}   & = &  - \frac{1}{6} ( D-F) \q , \q 
\kappa_{\Xi^0 \Xi^-}^K    =   \frac{1}{6} (D+5F) \q , \q 
\kappa_{\Xi^0 \Xi^-}^{\eta}    =   \frac{1}{6} (D+3F)  \no \\
\kappa_{p \Lambda}^{\pi}   & = &  - \frac{3}{2 \sqrt{6}} ( D+F) \q , \q 
\kappa_{p \Lambda}^K    =   - \frac{2}{\sqrt{6}} D \q , \q 
\kappa_{p \Lambda}^{\eta}    =   0  \no \\
\kappa_{\Lambda \Xi^- }^{\pi}   & = &  - \frac{1}{ \sqrt{6}} D \q , \q 
\kappa_{\Lambda \Xi^- }^K  = \frac{3}{2 \sqrt{6}} (D-F) \q , \q 
\kappa_{\Lambda \Xi^- }^{\eta}    =  \frac{1}{\sqrt{6}} D  \no \\
\kappa_{n \Sigma^-}^{\pi}   & = &  \frac{1}{3} ( D+F) \q , \q 
\kappa_{n \Sigma^-}^K    =    \frac{2}{3} F \q , \q 
\kappa_{n \Sigma^-}^{\eta}    =   - \frac{1}{6} ( D-3F)  \no \\
\kappa_{\Sigma^0 \Xi^-}^{\pi}   & = &   \frac{1}{3 \sqrt{2}} ( D+2F) \q , \q 
\kappa_{\Sigma^0 \Xi^-}^K  = \frac{1}{6\sqrt{2}} (7D + 5F) \q , \q 
\kappa_{\Sigma^0 \Xi^-}^{\eta}    =  \frac{1}{3\sqrt{2}} D  \no \\
\kappa_{\Sigma^+ \Xi^0}^{\phi} & = & \sqrt{2}\kappa_{\Sigma^0 \Xi^-}^{\phi}\qq.
\eeqa
\beqa
\rho_{pn}^{\pi}   & = &  \frac{4}{3} ( D+F) \q , \q 
\rho_{pn}^K    =   \frac{1}{2} D + \frac{1}{6} F \q , \q 
\rho_{pn}^{\eta}    =   0  \no \\
\rho_{\Lambda \Sigma^-}^{\pi}   & = &   \frac{1}{\sqrt{6}} ( D+2F) \q , \q 
\rho_{\Lambda \Sigma^-}^K  = \frac{1}{\sqrt{6}} ( D + F) \q , \q 
\rho_{\Lambda \Sigma^-}^{\eta}    =  0 \no \\
\rho_{\Xi^0 \Xi^-}^{\pi}   & = &  - \frac{1}{6} ( D-F) \q , \q 
\rho_{\Xi^0 \Xi^-}^K    =   \frac{1}{6} (D+5F) \q , \q 
\rho_{\Xi^0 \Xi^-}^{\eta}    =   \frac{1}{6} (D+3F)  \no \\
\rho_{p \Lambda}^{\pi}   & = &  - \frac{4}{\sqrt{6}} D \q , \q 
\rho_{p \Lambda}^K    =   \frac{1}{2\sqrt{6}} (D-3F) \q , \q 
\rho_{p \Lambda}^{\eta}    =   0  \no \\
\rho_{\Lambda \Xi^- }^{\pi}   & = &   \frac{3}{2 \sqrt{6}} (D-F) \q , \q 
\rho_{\Lambda \Xi^- }^K  = 0 \q , \q 
\rho_{\Lambda \Xi^- }^{\eta}    =  0  \no \\
\rho_{n \Sigma^-}^{\pi}   & = &  \frac{4}{3} F \q , \q 
\rho_{n \Sigma^-}^K    =    \frac{1}{6} (D+F) \q , \q 
\rho_{n \Sigma^-}^{\eta}    =   0  \no \\
\rho_{\Sigma^0 \Xi^-}^{\pi}   & = &   \frac{1}{3 \sqrt{2}} ( D-F) \q , \q 
\rho_{\Sigma^0 \Xi^-}^K  = \frac{4}{3\sqrt{2}} (D+F) \q , \q 
\rho_{\Sigma^0 \Xi^-}^{\eta}    =  \frac{1}{6\sqrt{2}} (D+3F)  \no \\
\rho_{\Sigma^+ \Xi^0}^{\phi} & = & \sqrt{2}\rho_{\Sigma^0 \Xi^-}^{\phi}\qq.
\eeqa
\beqa
\sigma_{pn}^{\pi} &=& \frac{10}{9}  \q , \q
\sigma_{pn}^{K} = \frac{2}{9}  \q , \q
\sigma_{pn}^{\eta} = 0  \no \\
\sigma_{\Lambda \Sigma^-}^{\pi} &=& \frac{2}{3 \sqrt{6}}  \q , \q
\sigma_{\Lambda \Sigma^-}^{K} = \frac{1}{3 \sqrt{6}}  \q , \q
\sigma_{\Lambda \Sigma^-}^{\eta} = 0  \no \\
\sigma_{\Xi^0 \Xi^-}^{\pi} &=& \frac{1}{18}  \q , \q
\sigma_{\Xi^0 \Xi^-}^{K} = - \frac{2}{9}  \q , \q
\sigma_{\Xi^0 \Xi^-}^{\eta} = - \frac{1}{6}  \no \\
\sigma_{p \Lambda}^{\pi} &=& - \frac{2}{\sqrt{6}}  \q , \q
\sigma_{p \Lambda}^{K} = - \frac{1}{\sqrt{6}}  \q , \q
\sigma_{p \Lambda}^{\eta} = 0  \no \\
\sigma_{\Lambda \Xi^-}^{\pi} &=& \frac{1}{\sqrt{6}}  \q , \q
\sigma_{\Lambda \Xi^-}^{K} = \frac{1}{\sqrt{6}}  \q , \q
\sigma_{\Lambda \Xi^-}^{\eta} = 0  \no \\
\sigma_{n \Sigma^-}^{\pi} &=& - \frac{2}{9}  \q , \q
\sigma_{n \Sigma^-}^{K} = - \frac{1}{9}  \q , \q
\sigma_{n \Sigma^-}^{\eta} =  0  \no \\
\sigma_{\Sigma^0 \Xi^-}^{\pi} &=& \frac{2}{9\sqrt{2}}  \q , \q
\sigma_{\Sigma^0 \Xi^-}^{K} = \frac{7}{9\sqrt{2}}  \q , \q
\sigma_{\Sigma^0 \Xi^-}^{\eta} = \frac{1}{3\sqrt{2}}  \no \\
\sigma_{\Sigma^0 \Xi^-}^{\phi} & = &\sqrt{2} \sigma_{\Sigma^0 \Xi^-}^{\phi}
\qq .  
\eeqa
\beqa
\epsilon_N^{\pi} & = & 1 \q , \q
\epsilon_N^{K}  =  \frac{1}{4} \q , \q
\epsilon_N^{\eta}  =  0 \q , \q
\epsilon_\Sigma^{\pi}  =  \frac{1}{6} \q , \q
\epsilon_\Sigma^{K}  =  \frac{5}{6} \q , \q
\epsilon_\Sigma^{\eta}  =  \frac{1}{4}  \no \\
\epsilon_\Lambda^{\pi} & = & \frac{3}{4} \q , \q
\epsilon_\Lambda^{K}  =  \frac{1}{2} \q , \q
\epsilon_\Lambda^{\eta}  =  0  \q , \q
\epsilon_\Xi^{\pi}  =  \frac{1}{4} \q , \q
\epsilon_\Xi^{K}  =  \frac{3}{4} \q , \q
\epsilon_\Xi^{\eta}  =  \frac{1}{4}  \qq .
\eeqa
There is an analogue formula for the form factor $G_3$ [Fig.~5]
\beqa
\delta G_3 [ij] (q^2) &=& - \frac{C^2}{\Lambda_\chi^2}\:
       \frac{ (M_i + M_j)^2 }{ m_{\tilde{\phi}}^2 - q^2} 
      \sum_{\phi=\pi,K,\eta} \: \bigg( \frac{4}{3} \kappa_{ij}^{\, \phi} 
         J^{\, \phi}( -\Delta,-\frac{q^2}{2 M_i}) +\frac{4}{3} 
       \rho_{ij}^{\, \phi} J^{\, \phi}(0,-\frac{q^2}{2 M_i}-\Delta) \no \\
& & \qq \q
  -\frac{10}{9}  H \:\sigma_{ij}^{\, \phi} 
     J^{\, \phi}(- \Delta,-\frac{q^2}{2 M_i}-\Delta)
   - \alpha_{ij} (\epsilon_i^{\, \phi}  + \epsilon_j^{\, \phi} )
      J^{\, \phi}(- \Delta,-\Delta)     \bigg)  
\eeqa
with $m_{\tilde{\phi}} = m_\pi$ 
for the decays $[ij] = [pn], [\Lambda\Sigma^-],
[\Xi^0\Xi^-]$ and $m_{\tilde{\phi}} = m_K$ otherwise.

\section{Results and discussion}
In this section we present the numerical results for the calculation
of the hadronic axial form factors.
We consider first the case with no resonances.
The values for our parameters are $F_{\pi}=93$ MeV,
$m_{\pi}=138$ MeV, $m_K=495$ MeV, and for the mass of the $\eta$ 
we use the GMO value for
the pseudoscalar mesons $m_{\eta}=566$ MeV.
The scale in dimensional regularization is chosen to be $\lambda= 1$ GeV.
The differences for $F_{\pi}$ and $m_{\eta}$ to  $\Fnod$--the pseudoscalar
decay constant in the chiral limit-- and to the physical mass of $\eta$,
respectively, appear only at higher orders.
We will restrict ourselves to these central values of the parameters
since a small variation in these parameters does only lead to some minor
changes in the results.

In baryon chiral perturbation theory, the transition between short and
long distance occurs around a distance scale of $\sim$1 fermi, or a momentum
scale of $\sim$200 MeV. This corresponds to the measured size of a baryon.
The effective field
theory treats the baryons and pions as point particles. This is
appropriate for the very long distance physics. 
However, for propagation at distances less then the separation
scale, the point particle theory is not an accurate representation of
the physics. The composite substructure becomes manifest below this
point.

Of course, the cutoff $\Lambda$ should not be taken so low in energy that it
removes any truly long distance physics. Also, while it can in principle be
taken much larger than the separation scale, this will lead to the
inclusion of spurious short distance physics which can upset the
convergence of the expansion. It is ideal to take the cutoff slightly
above the separation scale so that all of the long distance physics,
but little of the short distance physics, is included.
Therefore, we will vary the cutoff in the range
$\Lambda\geq 1/<r_B>\sim 300-600$ MeV.

The two unknown axial couplings $D$ and $F$ have to be fixed from 
phenomenology. We will choose the semileptonic decays
$n \rightarrow p$, $\Sigma^- \rightarrow \Lambda$, $\Lambda \rightarrow p$,
$\Xi^- \rightarrow \Lambda$, $\Sigma^- \rightarrow n$ and
$\Xi^- \rightarrow \Sigma^0$ to perform a least--squares fit
for $D$ and $F$.
In Table 1 the values for $D$ and $F$ for different values of the cutoff
$\Lambda$ are compared to the fit in dimensional regularization
and a fit at tree level.
At tree level the least--squares fit leads to
\beq
D = 0.80  \qq , \qq   F = 0.50  \q .
\eeq
Including the loop contributions,
one obtains in the cutoff scheme the values
\beq
D = 0.59 \pm 0.06  \qq , \qq   F = 0.36 \pm 0.05 \q ,
\eeq
whereas in the case of dimensional regularization a fit delivers
\beq
D = 0.44  \qq , \qq   F = 0.26  \q .
\eeq
In the latter case we neglected the analytic parts from the loops. 
The loop corrections lower in both cases the values for $D$ and $F$
but the change in $D$ and $F$ is larger in dimensional regularization.
On the other hand, 
the ratio $F/D$ remains the same in all cases: $F/D \simeq 0.61$.
The chiral expansions for $g_1$ at zero momentum transfer read
in the cutoff scheme for $\Lambda = 400$ MeV
\beqa
g_1[pn](0)    & = &  0.97 + 0.28  = 1.25  \qq (1.26) \no \\
g_1[\Lambda \Sigma^-](0)    & = &  0.49 + 0.17  = 0.66  \qq (0.62) \no \\
g_1[p \Lambda](0)    & = &  -0.69 - 0.26  = -0.95  \qq (-0.92) \no \\
g_1[\Lambda \Xi^-](0)    & = &  0.20 + 0.10  = 0.30  \qq (0.40) \no \\
g_1[n \Sigma^-](0)    & = &  0.23 + 0.07  = 0.30  \qq (0.39) \no \\
g_1[\Sigma^0 \Xi^-](0)    & = &  0.69 + 0.30  = 0.99  \qq (0.97) \no \\
g_1[\Xi^0 \Xi^-](0)    & = &  0.23 + 0.09  = 0.32  \qq . \no \\
\eeqa
Using dimensional regularization we obtain
\beqa
g_1[pn](0)    & = &  0.70 + 0.46  = 1.16  \qq (1.26) \no \\
g_1[\Lambda \Sigma^-](0)    & = &  0.36 + 0.28  = 0.64  \qq (0.62) \no \\
g_1[p \Lambda](0)    & = &  -0.50 - 0.48  = -0.98  \qq (-0.92) \no \\
g_1[\Lambda \Xi^-](0)    & = &  0.14 + 0.17  = 0.31  \qq (0.40) \no \\
g_1[n \Sigma^-](0)    & = &  0.18 + 0.14  = 0.32  \qq (0.39) \no \\
g_1[\Sigma^0 \Xi^-](0)    & = &  0.50 + 0.55  = 1.05  \qq (0.97) \no \\
g_1[\Xi^0 \Xi^-](0)    & = &  0.18 + 0.15  = 0.33  \qq . \no \\
\eeqa
The first number refers to the tree level contribution. The loop
contributions are summarized in the second number.
The numbers in the brackets are the experimental values. While the 
chiral series converge in the cutoff scheme, one cannot make a 
definite statement about the convergence in dimensional regularization.
Furthermore, we do not present the numerical results for the
decay $\Xi^0 \rightarrow \Sigma^+$ since it is related to 
$\Xi^- \rightarrow \Sigma^0$ by a factor of $\sqrt{2}$.
Also, the fit in dimensional regularization has
$\chi^2/d.o.f. = 2.8$, whereas we have $\chi^2/d.o.f. = 0.6$ in the
cutoff scheme for $\Lambda = 400$ MeV and
$\chi^2/d.o.f. = 1.0$ for the tree level fit. Note, that we have
increased the errors on the measurement of $g_A$ in the neutron decay
to 0.03 to avoid biasing the fit to the $D+F$ value favored by this decay. 
The values for the three axial form factors $g_1, g_2$ and $g_3$ at
zero momentum transfer in dimensional regularization and 
for different values of $\Lambda$ in the cutoff scheme 
are presented in Tab. 2.

Adding the decuplet, we set $\Delta = 231$ MeV, which is the average
octet--decuplet mass splitting, and the value of the coupling constant
$C$ is given by $C = 1.5$ from an overall fit to the decuplet
decays \cite{JM2}.
The introduction of the decuplet leads to an additional parameter $H$
which has to be fixed from phenomenology. We will determine $H$ along
with $D$ and $F$ by performing a least--squares fit to the form
factors $g_1(0)$.
One obtains in the cutoff scheme the values
\beq
D = 0.55 \pm 0.11  \qq , \qq   F = 0.46 \pm 0.05 \qq , \qq H = 3.0 \pm 5.0
\eeq
whereas in the case of dimensional regularization a fit delivers
\beq
D = 0.43  \qq , \qq   F = -0.14  \qq , \qq  H = -3.5  \q .
\eeq
It turns out that there are significant changes in the fit
in dimensional regularization
after including the decuplet. The values for $D$ and $F$ in the cutoff
scheme differ only slightly from the case without resonances.
No reliable estimate of the parameter $H$ can be given since the
uncertainty in the cutoff scheme is rather large and differs considerably
from the value in dimensional regularization.
The values of $D$, $F$ and $H$ for different values of $\Lambda$ are shown
in Tab.~3.
The chiral expansions in the cutoff scheme for $\Lambda = 400$ MeV read
\beqa
g_1[pn](0)    & = &  1.03 + 0.34 - 0.13  = 1.24  \qq (1.26) \no \\
g_1[\Lambda \Sigma^-](0)    & = &  0.48 + 0.19 - 0.02 = 0.65\qq (0.62) \no \\
g_1[p \Lambda](0)    & = &  -0.78 - 0.33 + 0.18 = -0.93  \qq (-0.92) \no \\
g_1[\Lambda \Xi^-](0)    & = &  0.30 + 0.17 - 0.15 = 0.32  \qq (0.40) \no \\
g_1[n \Sigma^-](0)    & = &  0.14 + 0.04 + 0.13  = 0.31  \qq (0.39) \no \\
g_1[\Sigma^0 \Xi^-](0)    & = &  0.73 + 0.36 - 0.08 = 1.01  \qq (0.97) \no \\
g_1[\Xi^0 \Xi^-](0)    & = &  0.14 + 0.04 + 0.09 = 0.27  \qq . \no \\
\eeqa
Using dimensional regularization we obtain
\beqa
g_1[pn](0)    & = &  0.29 + 0.21 + 0.70  = 1.20  \qq (1.26) \no \\
g_1[\Lambda \Sigma^-](0)    & = &  0.35 + 0.22 + 0.07= 0.64 \qq (0.62) \no \\
g_1[p \Lambda](0)    & = &  -0.01 - 0.02 - 0.80 = -0.83  \qq (-0.92) \no \\
g_1[\Lambda \Xi^-](0)    & = & -0.34 - 0.28 + 0.99 = 0.37  \qq (0.40) \no \\
g_1[n \Sigma^-](0)    & = &  0.56 + 0.51 - 0.73 = 0.34  \qq (0.39) \no \\
g_1[\Sigma^0 \Xi^-](0)    & = &  0.20 + 0.18 + 0.75 = 1.13  \qq (0.97) \no \\
g_1[\Xi^0 \Xi^-](0)    & = &  0.56 + 0.31 - 0.90  = -0.03  \qq . \no \\
\eeqa
The first and second number denote tree and loop contributions of
the baryon octet, respectively.
The third number is the loop contribution with intermediate decuplet states.
The $\chi^2/d.o.f.$ are $\chi^2/d.o.f. = 0.4$ and $\chi^2/d.o.f.=2.6$ for
the cutoff $\Lambda = 400$ MeV and dimensional regularization,
respectively.
The contributions from the resonance loops are well behaved in the cutoff 
scheme, whereas
they upset the behavior of the chiral series in dimensional regularization
and in most cases their contribution dominates.
A similar impact on the chiral series after the inclusion of the
decuplet was observed in the calculation of the baryon $\sigma$--terms
\cite{B}. 
Furthermore, while in the cutoff scheme we can predict
$g_1[\Xi^0 \Xi^-](0)     =   0.30 \pm 0.03$, the uncertainty in
dimensional regularization for this decay is large.
Setting the octet--decuplet mass splitting $\Delta = 0$ MeV
a least--squares fit delivers in the cutoff scheme
for $\Lambda = 400$ MeV
\beq
D = 0.57  \qq , \qq   F = 0.43  \qq , \qq  H = 1.5 
\eeq
whereas in the case of dimensional regularization we obtain
\beq
D = 0.48  \qq , \qq   F = 0.31  \qq , \qq  H = -1.4  \q .
\eeq
The latter result is in agreement with \cite{JM1} once one accounts
for the vanishing pion mass in that calculation.
Apparently, the fit for the three parameters in dimensional regularization
depends strongly on the value of the mass splitting $\Delta$ 
but in the cutoff scheme $D$ and $F$ are not altered significantly
by setting $\Delta = 0$ MeV. 
The reason for this is the large contribution of the decuplet
loops in dimensional regularization.
Again, there is a large uncertainty
in the parameter $H$. 
The values for the relativistic form factors $g_{1,2,3}$ can be found
in Tab.~4.
In the Figures 6 to 8 the $g_{1,2,3}$ are shown for small values of the
momentum transfer squared.
By neglecting the counterterms from the Lagrangian ${\it L}^{(2)}_{\phi B}$ 
one obtains similar results for the form factors $g_{1,3}$ but there is
a dramatic impact on $g_2$. While we have $|g_2/g_1| \simeq 0.3$
for most decays in our calculation, dropping these counterterms 
leads to $|g_2/g_1| \simeq 1$.

\section{Summary}
In this paper we have investigated the baryon axial currents both
in dimensional regularization and in a cutoff scheme.

\begin{enumerate}

\item[$\circ$] First, we presented the relation of the axial form factors
$G_{1,2,3}$
in heavy baryon chiral perturbation theory with the corresponding 
relativistic amplitudes $g_{1,2,3}$ in the convenient initial--baryon
restframe.
We calculated the chiral corrections to 
the axial currents by using both the lowest order effective
Lagrangian and their relativistic corrections at next order
in the heavy baryon formulation. The Goldstone boson integrals
are evaluated both in dimensional regularization and by using
a dipole regulator with a cutoff $\Lambda$ \cite{DHB}.
We have given the expressions for the form factors for general
momentum transfer.
With our definitions one of the form factors -- $G_2$ -- 
obtains contributions only from the relativistic corrections
of the meson baryon Lagrangian ${\it L}^{(2)}_{\phi B}$ of second
chiral order. Only these corrections, which have been neglected
in previous investigations, ensure the vanishing of $g_2[pn]$ at tree level.
Otherwise, one would obtain $g_2[pn] = -g_1[pn] \simeq -1.26$.

The cutoff parameter induces an additional mass scale that does not vanish
in the chiral limit and, therefore, destroys the strict chiral counting
scheme. We are able to show that to the order we are working the
physics does not depend on $\Lambda$, since one is able to absorb
the effects of $\Lambda$ into a renormalization of the coupling
constants.

\item[$\circ$] The spin--3/2 decuplet is separated from the octet
by $\Delta =231$ MeV in average which is smaller than the kaon or eta mass.
Therefore, we proceeded by adding the decuplet to the effective theory.
The appearance of the mass scale $\Delta$ destroys in the case of
dimensional regularization the one--to--one correspondence between
meson loops and the expansion in small momenta and quark masses.
No further complications arise in the cutoff scheme since the strict chiral
counting scheme has already been spoilt by introducing the scale $\Lambda$.

\item[$\circ$] We performed a least--squares fit to the semileptonic
hyperon decays. The values for the two axial couplings $D$ and $F$
are at tree level $D = 0.80$ and $F = 0.50$.
Including the chiral corrections from the loops without resonances
we obtain $D = 0.44$, $F = 0.26$ in dimensional regularization
and $D = 0.59 \pm 0.06$, $F = 0.36 \pm 0.05$ by using a cutoff.
The uncertainty in the fit stems from the variation of the cutoff
parameter. In our analysis the parameter $\Lambda$ ranges from 300 to 600 MeV
to account for all the long distance physics, but little of the
short distance physics, which are not described appropriately by
the effective theory, is included.
The results are in good agreement with the experimental data
and we have $\chi^2/d.o.f. = 0.6$ for the cutoff $\Lambda= 400$ MeV
to be compared with $\chi^2/d.o.f. = 1.0$ at tree level.
In dimensional regularization, one obtains $\chi^2/d.o.f. = 2.8$.
While the chiral expansions of the form factors converge in the cutoff
scheme, one cannot make a definit statement about the convergence
in dimensional regularization.
After fixing $D$ and $F$ from experiment, results for the three
momentum dependent axial form factors are presented.

The introduction of the decuplet leads to an additional parameter $H$
which has to be fixed from phenomenology. 
Adding the loop contributions with intermediate decuplet states
alter the results significantly in dimensional regularization .
A least--squares fit to the semileptonic decays delivers the values
$D = 0.43$, $F = -0.14$ and $H= -3.5$.
The decuplet contributions tend to dominate the pieces from tree level
and loops involving only baryon octet fields.
In the cutoff scheme the contributions from the resonances behave moderate
and one obtains $D = 0.55 \pm 0.11$, $F = 0.46 \pm 0.05$ and 
$H= 3.0 \pm 5.0$.
No reliable estimate of the parameter $H$ can be given since the
uncertainty in the cutoff scheme is rather large and differs considerably
from the value in dimensional regularization.

Setting the average octet--decuplet mass splitting $\Delta =0$ MeV 
leads to $D = 0.57$, $F = 0.43$, $H= 1.5$ and 
$D = 0.48$, $F = 0.31$, $H= -1.4$ using a cutoff $\Lambda = 400$ MeV
and in dimensional 
regularization, respectively. The latter case is in agreement with
\cite{JM1}, once one accounts for the vanishing pion mass in that work.

\end{enumerate}

\section*{Acknowledgements}
This work was supported in part by the Deutsche Forschungsgemeinschaft.

\appendix
\def\theequation{\Alph{section}.\arabic{equation}}
\setcounter{equation}{0}
\section{Integrals}
The fundamental integral in the calculation of the axial form factors is
in the cutoff scheme
\beqa  
J^{\, \phi} (x,y) &=&  \frac{4}{3 \pi^2}
\int   d^4 l  \, \, \frac{ i^3 \, ( S \cdot l )^2}{
       [l^2 - m_{\phi}^2 + i \epsilon] \, [ v \cdot l +x + i \epsilon ] 
       \, [ v \cdot l + y + i \epsilon ] }
       \bigg( \frac{ \Lambda^2}{\Lambda^2 - l^2} \bigg)^2  \no \\
  &=&  \:  - \frac{1}{3} {\Lambda^4\over (\Lambda^2-m_\phi^2)^2}\bigg(
     \Lambda^2-m_\phi^2 + m_\phi^2 \ln {m_\phi^2\over\Lambda^2} \bigg) \no \\
  & &  \: + \frac{2}{3}\frac{1}{x-y}
        {\Lambda^4\over (\Lambda^2-m_\phi^2)^2}\bigg(
      \Big[ \, y (m_\phi^2 - y^2) - x (m_\phi^2 - x^2) \Big] \ln
              {m_\phi^2\over\Lambda^2}  \no \\
  & & \: + \Big[ \frac{3}{2} m_\phi^2 - x^2 - \frac{1}{2} \Lambda^2 \Big]
           \, f(\Lambda,x) - \Big[ m_\phi^2 - x^2 \Big] \, f(m_\phi,x) \no \\
  & & \: - \Big[ \frac{3}{2} m_\phi^2 - y^2 - \frac{1}{2} \Lambda^2 \Big]
           \, f(\Lambda,y) + \Big[ m_\phi^2 - y^2 \Big] \, f(m_\phi,y) \bigg)
\eeqa
with
\beqa
f(u,v)  & = & 2 \,\sqrt{ u^2 - v^2} \,\arccos \frac{-v}{u} \qq ; 
                  \qq \mbox{for} \q |u| > |v|  \no \\
f(u,v)  & = & 2 \,\sqrt{ v^2 - u^2} \,\ln \Big[ \,\frac{v}{u} + 
                 \sqrt{ \frac{v^2}{u^2} -1\,} \Big]  \qq ; 
                  \qq \mbox{for} \q \frac{v}{u} > 1   \no \\ 
f(u,v)  & = & - 2 \,\sqrt{ v^2 - u^2} \,\ln \Big[ \,-\frac{v}{u} + 
                 \sqrt{ \frac{v^2}{u^2} -1\,} \Big]  \qq ; 
                  \qq \mbox{for} \q \frac{v}{u} < - 1   \qq .
\eeqa
In dimensional regularization the nonanalytic part of the integral reads
\beqa
J_{dim}^{\, \phi} (x,y) &=&  \frac{64 \pi^2}{3}
\int   \frac{d^d l}{(2 \pi)^d}   \, \, \frac{ i^3 \, ( S \cdot l )^2}{
       [l^2 - m_{\phi}^2 + i \epsilon] \, [ v \cdot l +x + i \epsilon ] 
       \, [ v \cdot l + y + i \epsilon ] } \no\\
& = & \: -  \, \bigg(
           m_\phi^2 -\frac{2}{3} \Big[ x^2 + xy + y^2 \Big] \bigg) 
           \ln \frac{m_\phi^2}{\lambda^2} \no \\
& & \: + \frac{2}{3} \frac{x^2-m_\phi^2}{x-y} \, f(m_\phi,x) \:
    \: + \frac{2}{3} \frac{m_\phi^2-y^2}{x-y} \, f(m_\phi,y) \qq ,
\eeqa
with $\lambda$ the scale introduced in dimensional regularization. \\
In Section 2 we use for the loop integrals involving only the baryon
fields the notation
\beq
I^{\, \phi}(x) = - J^{\, \phi}(0,x) \qq .
\eeq

\newpage

\section*{Table captions}

\begin{enumerate}

\item[Table 1] Given are the values for the couplings $D$ and $F$ 
               and the pertinent $\chi^2/d.o.f.$
               from a least--squares fit to the axial form factors $g_1(0)$
               both in dimensional regularization and for various values
               of the cutoff $\Lambda$ in MeV.

\item[Table 2] Given are the values for the three axial form factors
               $g_{1,2,3}$ at zero momentum transfer $q^2 = 0$
               both in dimensional regularization and for different values
               of the cutoff $\Lambda$ in MeV.

\item[Table 3] Given are the values for the couplings $D$, $F$ and $H$ 
               and the pertinent $\chi^2/d.o.f.$
               from a least--squares fit to the axial form factors $g_1(0)$
               including contributions from the spin--3/2 decuplet
               both in dimensional regularization and for various values
               of the cutoff $\Lambda$ in MeV.

\item[Table 4] Given are the values for the three axial form factors
               $g_{1,2,3}$ at zero momentum transfer $q^2 = 0$
               including contributions from the spin--3/2 decuplet
               both in dimensional regularization and for different values
               of the cutoff $\Lambda$ in MeV.

\end{enumerate}

\vskip 1.2in
%%%% figure captions

\section*{Figure captions}

\begin{enumerate}

\item[Fig.1] Tree graphs to the form factors $G_{1,2}$ and $G_3$.
             Solid and dashed lines denote octet baryons and Goldstone
             bosons, respectively. 
             The solid circle denotes a strong vertex and the solid
             square represents the axial current.

\item[Fig.2] Loop contributions to $G_1$.
             Solid and dashed lines denote octet baryons and Goldstone
             bosons, respectively. 
             The solid circle denotes a strong vertex and the solid
             square represents the axial current.

\item[Fig.3] Loop contributions to $G_3$.
             Solid and dashed lines denote octet baryons and Goldstone
             bosons, respectively. 
             The solid circle denotes a strong vertex and the solid
             square represents the axial current.

\item[Fig.4] Loop contributions to $G_1$ with intermediate decuplet states.
             Solid and dashed lines denote octet baryons and Goldstone
             bosons, respectively. The double line represents the decuplet.
             The solid circle denotes a strong vertex and the solid
             square represents the axial current. 

\item[Fig.5] Loop contributions to $G_3$ with intermediate decuplet states. 
             Solid and dashed lines denote octet baryons and Goldstone
             bosons, respectively. The double line represents the decuplet.
             The solid circle denotes a strong vertex and the solid
             square represents the axial current. 

\item[Fig.6] The results for the form factor $g_1(t)$ are presented:
             a) in the cutoff scheme with $\Lambda= 400$ MeV and ground state
             octet bayons only;  b) in the cutoff scheme with $\Lambda= 400$
             MeV and including the decuplet; c) in dimensional
             regularization and ground state octet baryons only;
             d) in dimensional regularization including the decuplet.
             The different lines refer to the following decays:
             $n \rightarrow p$: continous line; $\Lambda \rightarrow p$:
             dot--dashed; $\Sigma^- \rightarrow \Lambda$: broken line;
             $\Sigma^- \rightarrow n$: dashed; $\Xi^- \rightarrow \Lambda$:
             dotted; $\Xi^- \rightarrow \Sigma^0$: dot--dot--dashed;
             $\Xi^- \rightarrow \Xi^0$: dot--dash--dashed.

\item[Fig.7] The results for the form factor $g_2(t)$ are presented:
             a) in the cutoff scheme with $\Lambda= 400$ MeV and ground state
             octet bayons only;  b) in the cutoff scheme with $\Lambda= 400$
             MeV and including the decuplet; c) in dimensional
             regularization and ground state octet baryons only;
             d) in dimensional regularization including the decuplet.
             The different lines refer to the following decays:
             $n \rightarrow p$: continous line; $\Lambda \rightarrow p$:
             dot--dashed; $\Sigma^- \rightarrow \Lambda$: broken line;
             $\Sigma^- \rightarrow n$: dashed; $\Xi^- \rightarrow \Lambda$:
             dotted; $\Xi^- \rightarrow \Sigma^0$: dot--dot--dashed;
             $\Xi^- \rightarrow \Xi^0$: dot--dash--dashed.

\item[Fig.8] The results for the form factor $g_3(t)$ are presented:
             a) in the cutoff scheme with $\Lambda= 400$ MeV and ground state
             octet bayons only;  b) in the cutoff scheme with $\Lambda= 400$
             MeV and including the decuplet; c) in dimensional
             regularization and ground state octet baryons only;
             d) in dimensional regularization including the decuplet.
             The different lines refer to the following decays:
             $n \rightarrow p$: continous line; $\Lambda \rightarrow p$:
             dot--dashed; $\Sigma^- \rightarrow \Lambda$: broken line;
             $\Sigma^- \rightarrow n$: dashed; $\Xi^- \rightarrow \Lambda$:
             dotted; $\Xi^- \rightarrow \Sigma^0$: dot--dot--dashed;
             $\Xi^- \rightarrow \Xi^0$: dot--dash--dashed.

\end{enumerate}

\newpage

%%%%%%% Tables   %%%%%%
\begin{center}

\begin{table}[bht]  \label{tab1}
\begin{center}
 % \medskip 
\begin{tabular}{l|c|c|c|c|c}
  & dim. & $\Lambda=300$&$\Lambda=400$&$\Lambda=500$&$\Lambda=600$\\
\hline
$ D $  & 0.44 & 0.64 & 0.60 & 0.56 & 0.54 \\
$ F $  & 0.26 & 0.40 & 0.37 & 0.35 & 0.33 \\
$ \chi^2/d.o.f. $  & 2.8 & 0.5 & 0.6 & 0.8 & 1.0 \\
\hline
\end{tabular}
\end{center}
\end{table}
\vskip 0.7cm

Table  1

\vskip 1.5cm

\begin{table}[bht]  \label{tab2}
\begin{center}
 % \medskip 
\begin{tabular}{l|c|c|c|c|c}
  & dim. & $\Lambda=300$&$\Lambda=400$&$\Lambda=500$&$\Lambda=600$\\
\hline
$g_1 \, [pn] \, (0)  $  & 1.16 & 1.27 & 1.25 & 1.23 & 1.21 \\
$g_1 \, [\Lambda \Sigma^-] \, (0)  $  & 0.64 & 0.66 & 0.66 & 0.66 & 0.65 \\
$g_1 \, [\Xi^0 \Xi^-] \, (0)  $  & 0.33 & 0.31 & 0.32 & 0.32 & 0.32 \\
$g_1 \, [p \Lambda ] \, (0)  $  & -0.98 & -0.95 & -0.95 & -0.96 & -0.96 \\
$g_1 \, [\Lambda \Xi^- ] \, (0)  $  & 0.31 & 0.30 & 0.30 & 0.30 & 0.31 \\
$g_1 \, [n \Sigma^-] \, (0)  $  & 0.32 & 0.30 & 0.30 & 0.30 & 0.31 \\
$g_1 \, [\Sigma^0 \Xi^-] \, (0)  $  & 1.05 & 0.97 & 0.99 & 1.00 & 1.01 \\
\hline
$g_2 \, [pn] \, (0)  $  & -0.12 &-0.23 &-0.28 &-0.32 &-0.35 \\
$g_2 \, [\Lambda \Sigma^-] \, (0)  $  &-0.14 &-0.16 &-0.20 &-0.22 &-0.25 \\
$g_2 \, [\Xi^0 \Xi^-] \, (0)  $  & 0.71 & 0.72 & 0.65 & 0.59 & 0.54 \\
$g_2 \, [p \Lambda ] \, (0)  $  & 0.34 & 0.31 & 0.37 & 0.42 & 0.46 \\
$g_2 \, [\Lambda \Xi^- ] \, (0)  $  &-0.12 &-0.11 &-0.13 &-0.15 &-0.16 \\
$g_2 \, [n \Sigma^-] \, (0)  $  &-0.12 &-0.10 &-0.12 &-0.14 &-0.15 \\
$g_2 \, [\Sigma^0 \Xi^-] \, (0)  $  &-0.39 &-0.30 &-0.37 &-0.43 &-0.48 \\
\hline
$g_3 \, [pn] \, (0)  $  & -215.9 & -236.0 & -232.3 & -228.9 & -226.0 \\
$g_3 \, [\Lambda \Sigma^-] \, (0)  $  & -174.7 & -179.5 & -179.5 & -179.1 &
                                     -178.5  \\
$g_3 \, [\Xi^0 \Xi^-] \, (0)  $  & -118.4 & -113.6 & -115.5 & -116.9 & 
                                    -117.9 \\
$g_3 \, [p \Lambda ] \, (0)  $  & 16.2 & 15.6 & 15.7 & 15.9 & 16.0 \\
$g_3 \, [\Lambda \Xi^- ] \, (0)  $  & -7.2 & -6.8 & -7.0 & -7.0 & -7.1 \\
$g_3 \, [n \Sigma^-] \, (0)  $  & -5.6 & -5.1 & -5.2 & -5.3 & -5.4 \\
$g_3 \, [\Sigma^0 \Xi^-] \, (0)  $  & -26.5 & -24.2 & -24.8 & -25.2 &-25.6 \\
\hline
\end{tabular}
\end{center}
\end{table}
\vskip 0.7cm

Table  2

\newpage

\begin{table}[tbh]  \label{tab3}
\begin{center}
 % \medskip 
\begin{tabular}{l|c|c|c|c|c}
  & dim. & $\Lambda=300$&$\Lambda=400$&$\Lambda=500$&$\Lambda=600$\\
\hline
$ D $  & 0.43 & 0.66 & 0.59 & 0.52 & 0.45 \\
$ F $  & -0.14 & 0.48 & 0.44 & 0.43 & 0.46 \\
$ H $  & -3.5 & 9.1 & 3.4 & 1.9 & 1.4 \\
$ \chi^2/d.o.f. $  & 2.6 & 0.4 & 0.4 & 0.3 & 0.3 \\
\hline
\end{tabular}
\end{center}
\end{table}
\vskip 0.7cm

Table  3

\vskip 1.5cm

\begin{table}[tbh]  \label{tab4}
\begin{center}
 % \medskip 
\begin{tabular}{l|c|c|c|c|c}
  & dim. & $\Lambda=300$&$\Lambda=400$&$\Lambda=500$&$\Lambda=600$\\
\hline
$g_1 \, [pn] \, (0)  $  & 1.20 & 1.25 & 1.24 & 1.24 & 1.23 \\
$g_1 \, [\Lambda \Sigma^-] \, (0)  $  & 0.64 & 0.65 & 0.65 & 0.65 & 0.65 \\
$g_1 \, [\Xi^0 \Xi^-] \, (0)  $  & -0.03 & 0.28 & 0.27 & 0.26 & 0.23 \\
$g_1 \, [p \Lambda ] \, (0)  $  & -0.83 & -0.94 & -0.93 & -0.93 & -0.91 \\
$g_1 \, [\Lambda \Xi^- ] \, (0)  $  & 0.37 & 0.31 & 0.32 & 0.33 & 0.36 \\
$g_1 \, [n \Sigma^-] \, (0)  $  & 0.34 & 0.31 & 0.31 & 0.31 & 0.32 \\
$g_1 \, [\Sigma^0 \Xi^-] \, (0)  $  & 1.13 & 1.00 & 1.01 & 1.02 & 1.02 \\
\hline
$g_2 \, [pn] \, (0)  $  &-0.33 &-0.21 &-0.28 &-0.33 &-0.37 \\
$g_2 \, [\Lambda \Sigma^-] \, (0)  $  &-0.23 &-0.16 &-0.19 &-0.21 &-0.24 \\
$g_2 \, [\Xi^0 \Xi^-] \, (0)  $  & 0.89 & 0.76 & 0.69 & 0.65 & 0.63 \\
$g_2 \, [p \Lambda ] \, (0)  $  & 0.30 & 0.30 & 0.35 & 0.38 & 0.40 \\
$g_2 \, [\Lambda \Xi^- ] \, (0)  $  &-0.24 &-0.13 &-0.15 &-0.19 &-0.23 \\
$g_2 \, [n \Sigma^-] \, (0)  $  &-0.20 &-0.11 &-0.13 &-0.15 &-0.17 \\
$g_2 \, [\Sigma^0 \Xi^-] \, (0)  $  &-0.60 &-0.34 &-0.40 &-0.45 &-0.48 \\
\hline
$g_3 \, [pn] \, (0)  $  & -222.9 & -231.8 & -231.2 & -230.5 & -230.0 \\
$g_3 \, [\Lambda \Sigma^-] \, (0)  $  & -175.9 & -178.1 & -177.2 & -176.5 &
                                       -176.9  \\
$g_3 \, [\Xi^0 \Xi^-] \, (0)  $  & 11.2 & -100.6 & -98.6 & -93.5 & -84.3 \\
$g_3 \, [p \Lambda ] \, (0)  $  & 13.9 & 15.3 & 15.3 & 15.3 & 15.1 \\
$g_3 \, [\Lambda \Xi^- ] \, (0)  $  & -9.1 & -7.1 & -7.4 & -7.7 & -8.3 \\
$g_3 \, [n \Sigma^-] \, (0)  $  & -5.7 & -5.3 & -5.4 & -5.6 & -5.9 \\
$g_3 \, [\Sigma^0 \Xi^-] \, (0)  $  & -28.6 & -25.1 & -25.4 & -25.6 &-25.8 \\
\hline
\end{tabular}
\end{center}
\end{table}
\vskip 0.7cm

Table  4

\vskip 1.5cm

\end{center}

\newpage

%%%%%%% Figures   %%%%%%
\begin{center}
 
\begin{figure}[bth]
\centering
%\leavemode
\centerline{
\epsfbox{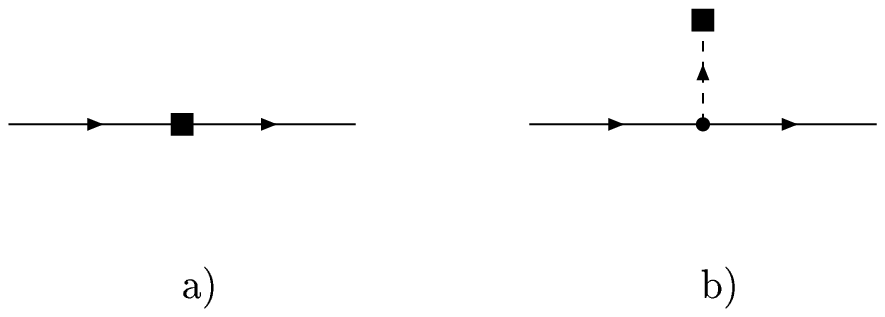}}
\end{figure}

\vskip 0.4cm

Figure 1

\vskip 1.9cm

\begin{figure}[tbh]
\centering
%\leavemode
\centerline{
\epsfbox{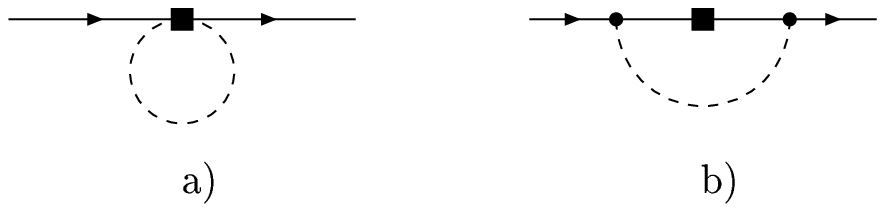}}
\end{figure}

\vskip 0.4cm

Figure 2

\vskip 1.9cm

\begin{figure}[tbh]
\centering
%\leavemode
\centerline{
\epsfbox{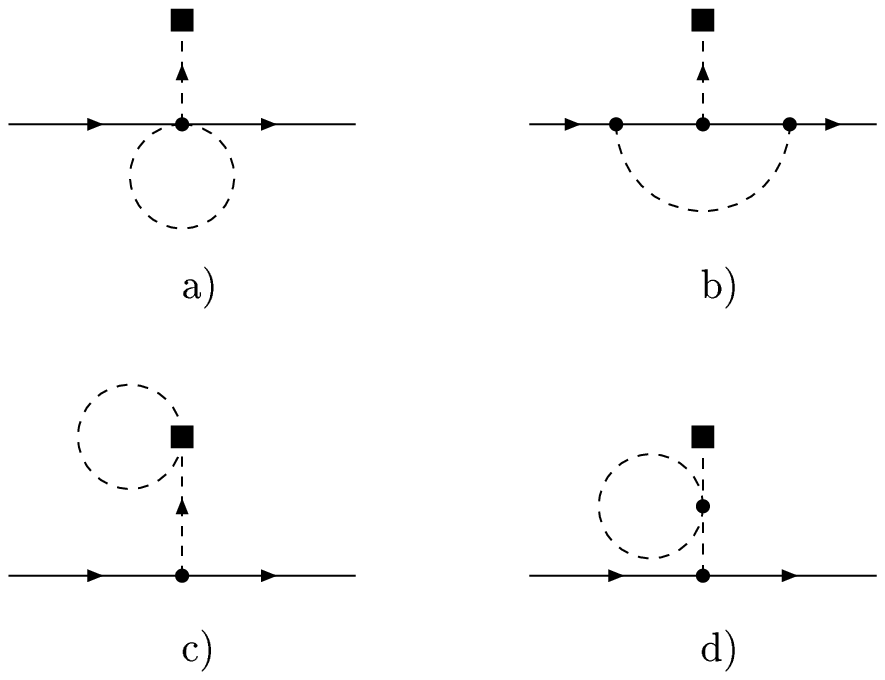}}
\end{figure}

\vskip 0.4cm

Figure 3

\vskip 1.9cm

\clearpage

\begin{figure}[tbh]
\centering
%\leavemode
\centerline{
\epsfbox{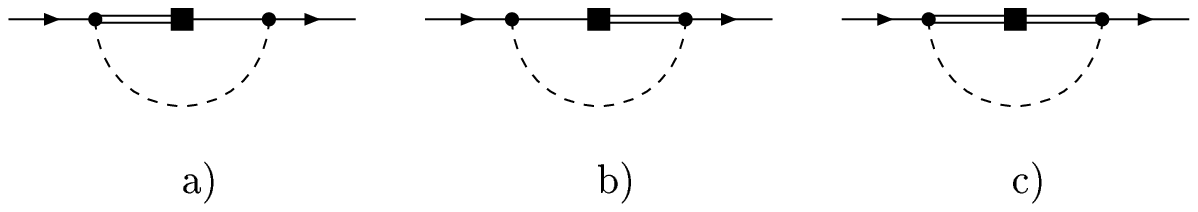}}
\end{figure}

\vskip 0.4cm

Figure 4

\vskip 1.9cm

\begin{figure}[tbh]
\centering
%\leavemode
\centerline{
\epsfbox{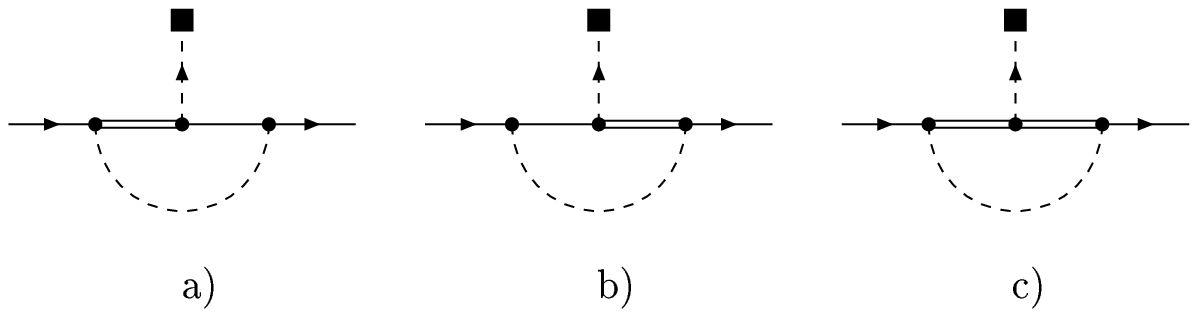}}
\end{figure}

\vskip 0.4cm

Figure 5

\begin{figure}[tbh]
\centering
%\leavemode
\begin{picture}(300,380)  
\put(0,260){\makebox(100,120){\epsfig{file=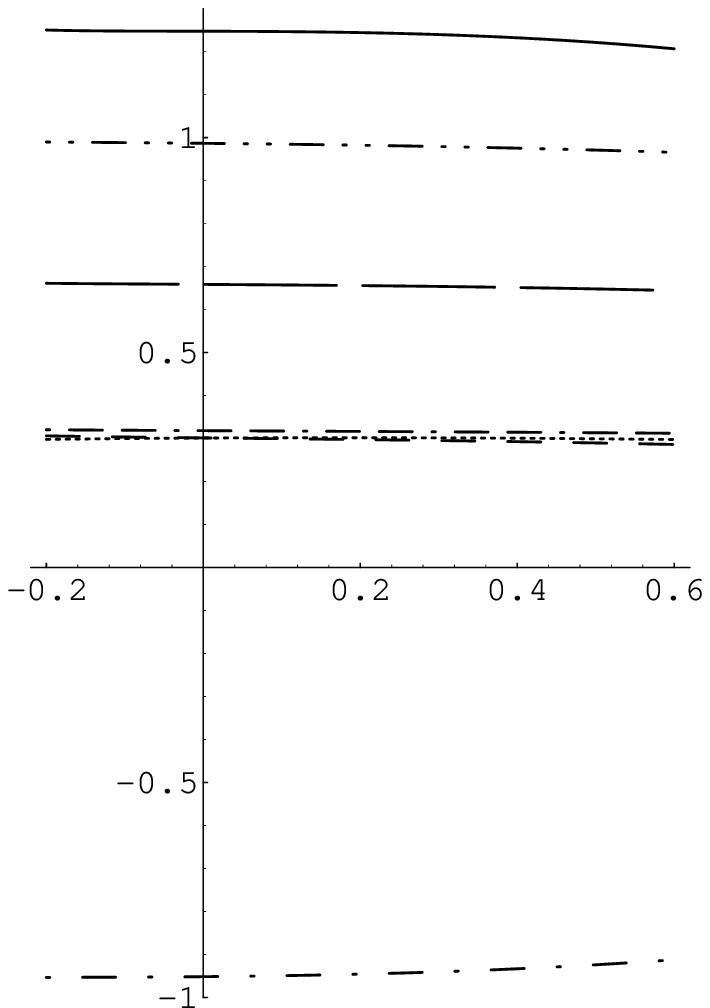,width=7.0cm}}}
\put(200,260){\makebox(100,120){\epsfig{file=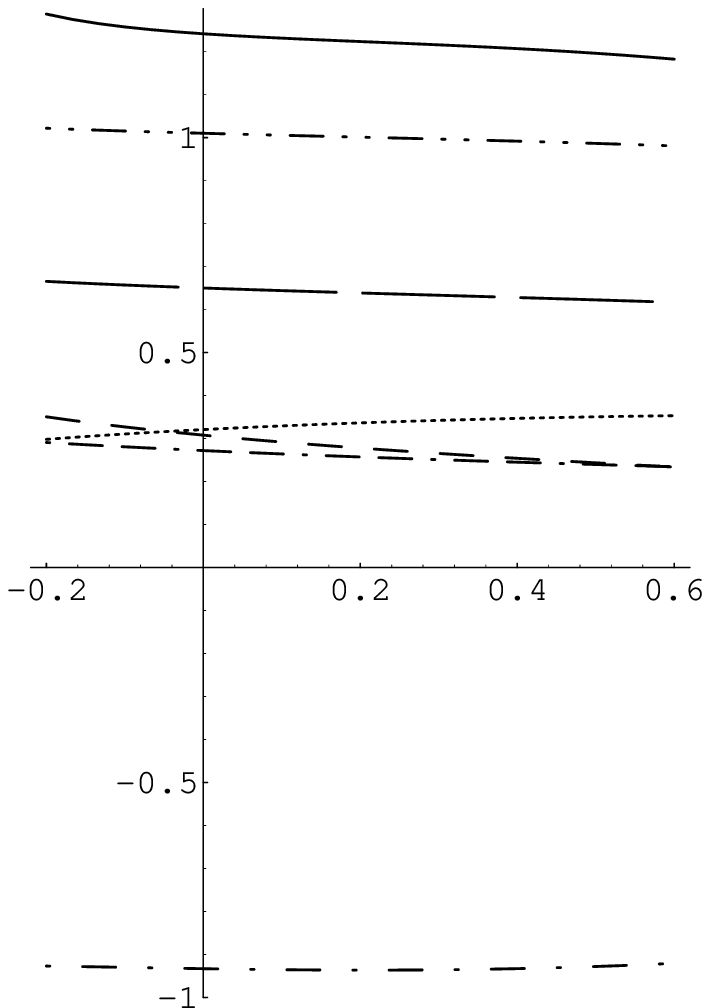,width=7.0cm}}}
\put(0,0){\makebox(100,120){\epsfig{file=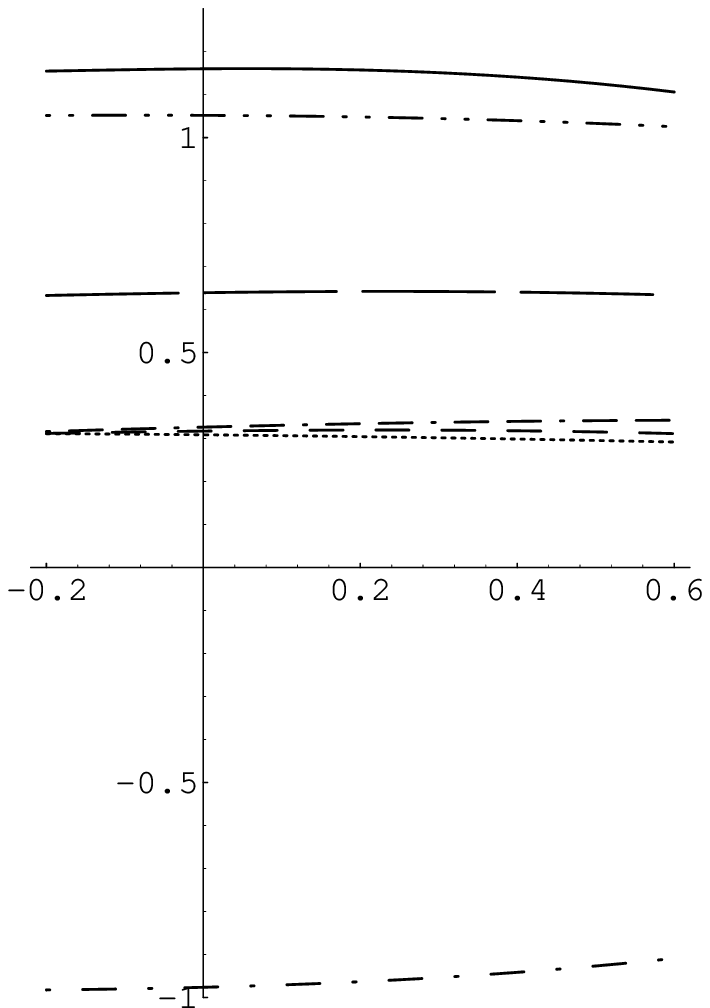,width=7.0cm}}}
\put(200,0){\makebox(100,120){\epsfig{file=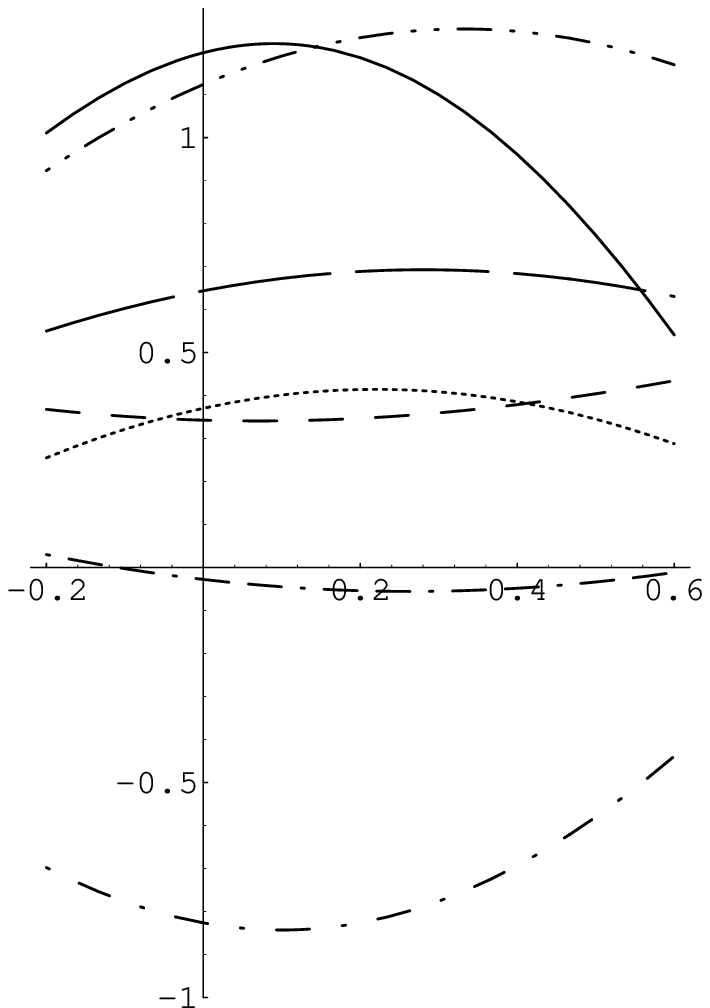,width=7.0cm}}}
\put(50,200){$a)$}
\put(250,200){$b)$}
\put(50,-60){$c)$}
\put(250,-60){$d)$}
\put(110,292){\scriptsize{$q^2 \, [GeV^2]$}}
\put(310,292){\scriptsize{$q^2 \, [GeV^2]$}}
\put(110,32){\scriptsize{$q^2 \, [GeV^2]$}}
\put(310,32){\scriptsize{$q^2 \, [GeV^2]$}}
\end{picture}
\vskip 2.8cm

Figure 6

\end{figure}

\begin{figure}[tbh]
\centering
%\leavemode
\begin{picture}(300,380)  
\put(0,260){\makebox(100,120){\epsfig{file=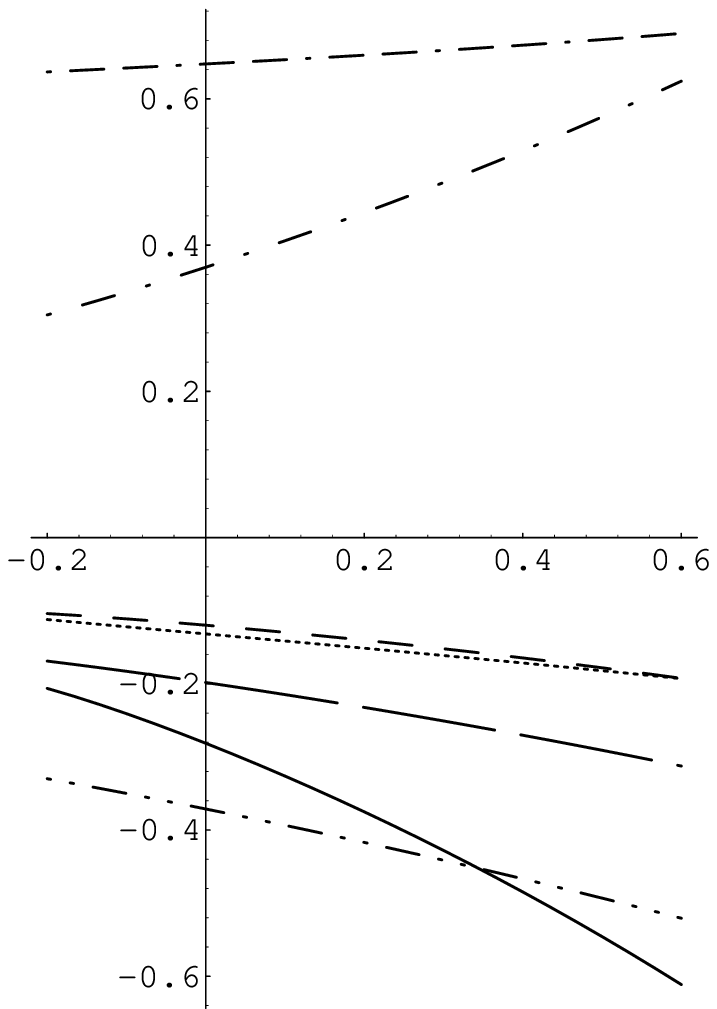,width=7.0cm}}}
\put(200,260){\makebox(100,120){\epsfig{file=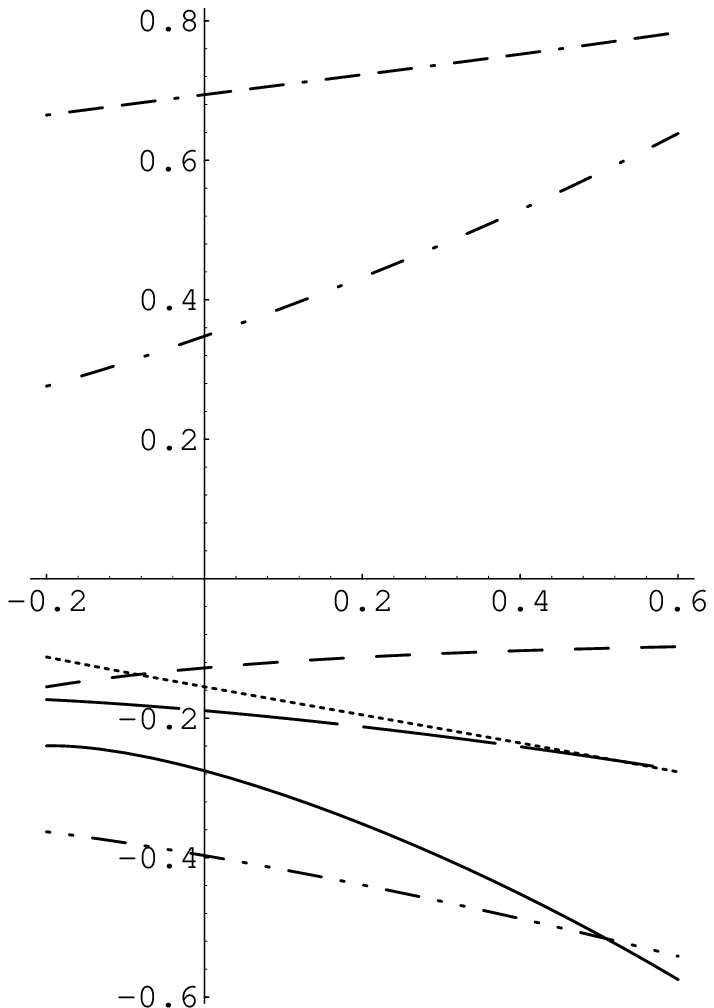,width=7.0cm}}}
\put(0,0){\makebox(100,120){\epsfig{file=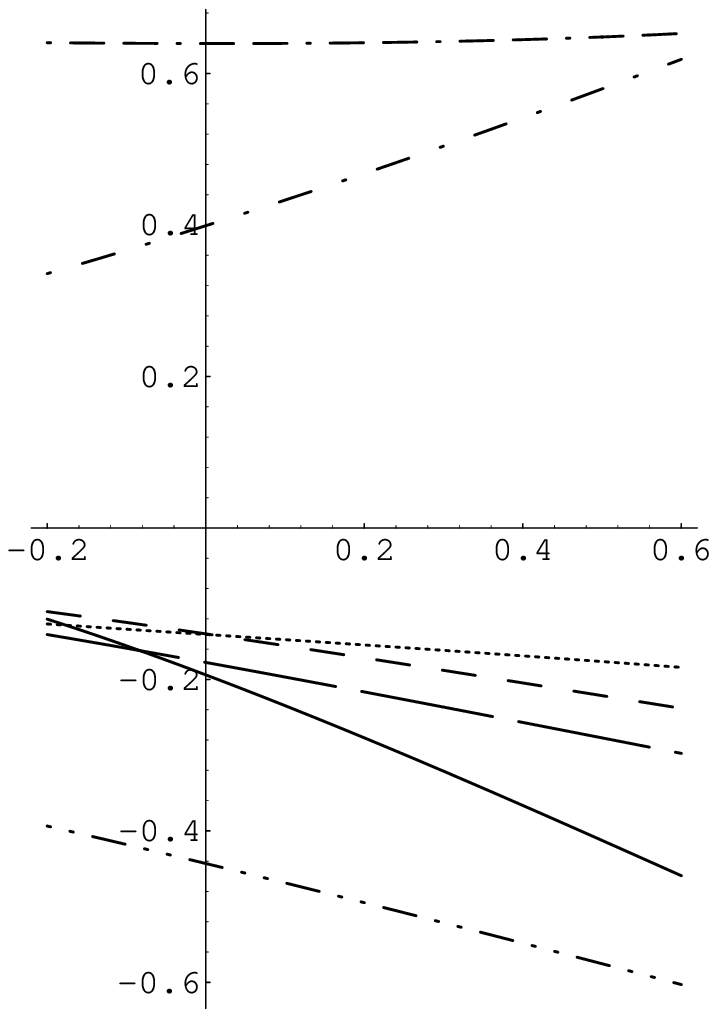,width=7.0cm}}}
\put(200,0){\makebox(100,120){\epsfig{file=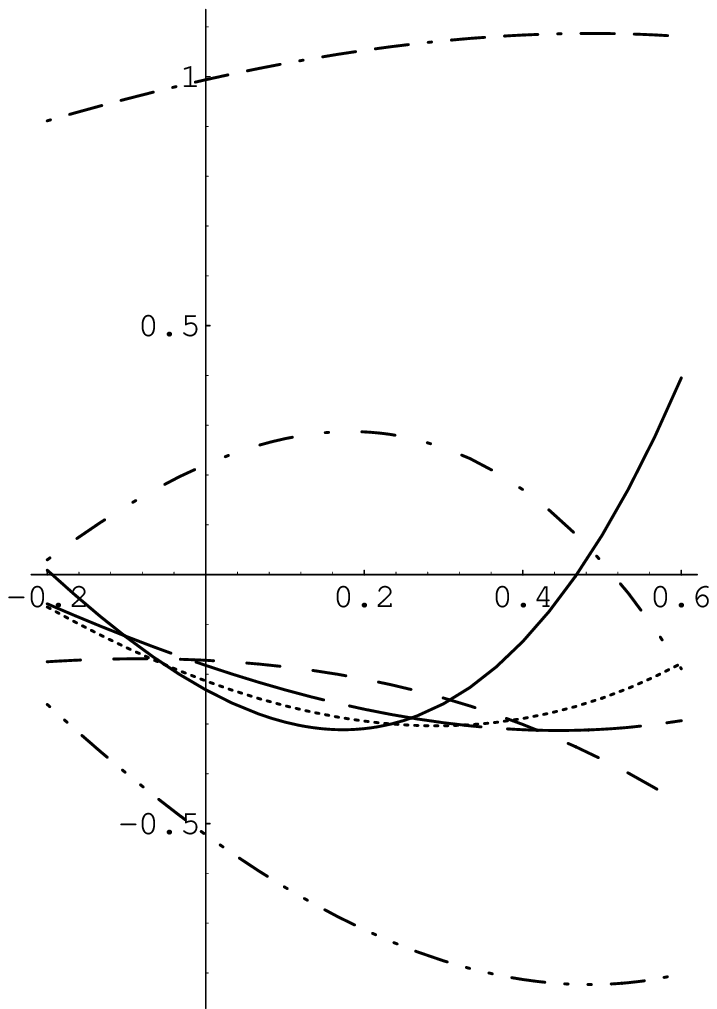,width=7.0cm}}}
\put(50,200){$a)$}
\put(250,200){$b)$}
\put(50,-60){$c)$}
\put(250,-60){$d)$}
\put(115,299){\scriptsize{$q^2 \, [GeV^2]$}}
\put(317,292){\scriptsize{$q^2 \, [GeV^2]$}}
\put(115,40){\scriptsize{$q^2 \, [GeV^2]$}}
\put(317,32){\scriptsize{$q^2 \, [GeV^2]$}}
\end{picture}
\vskip 2.8cm

Figure 7

\end{figure}

\begin{figure}[tbh]
\centering
%\leavemode
\begin{picture}(300,380)  
\put(0,260){\makebox(100,120){\epsfig{file=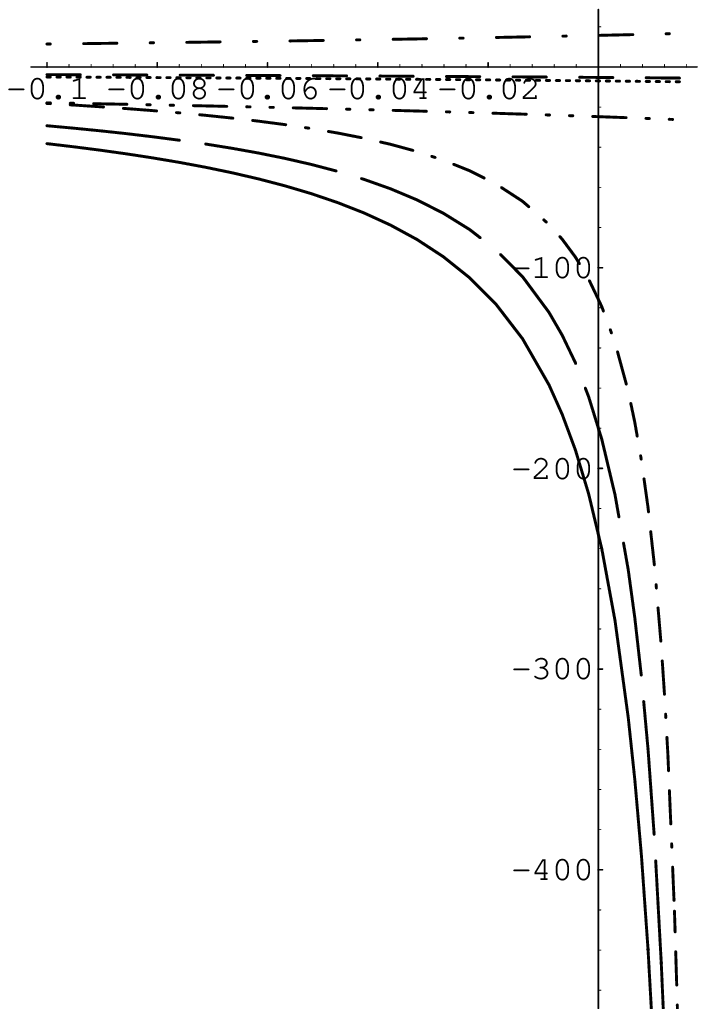,width=7.0cm}}}
\put(200,260){\makebox(100,120){\epsfig{file=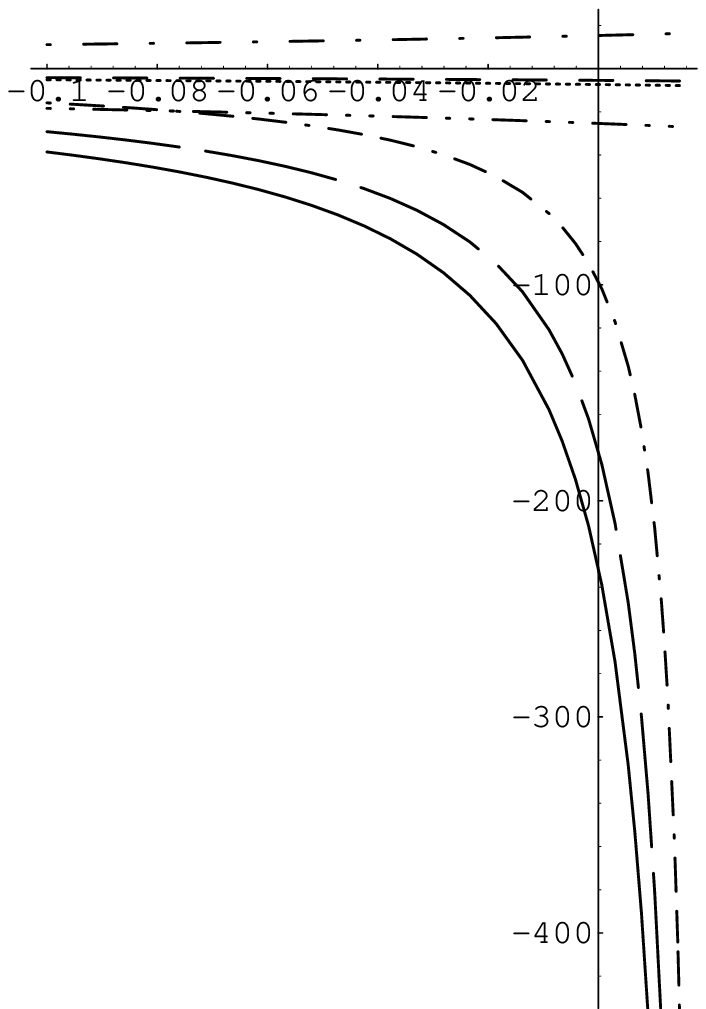,width=7.0cm}}}
\put(0,0){\makebox(100,120){\epsfig{file=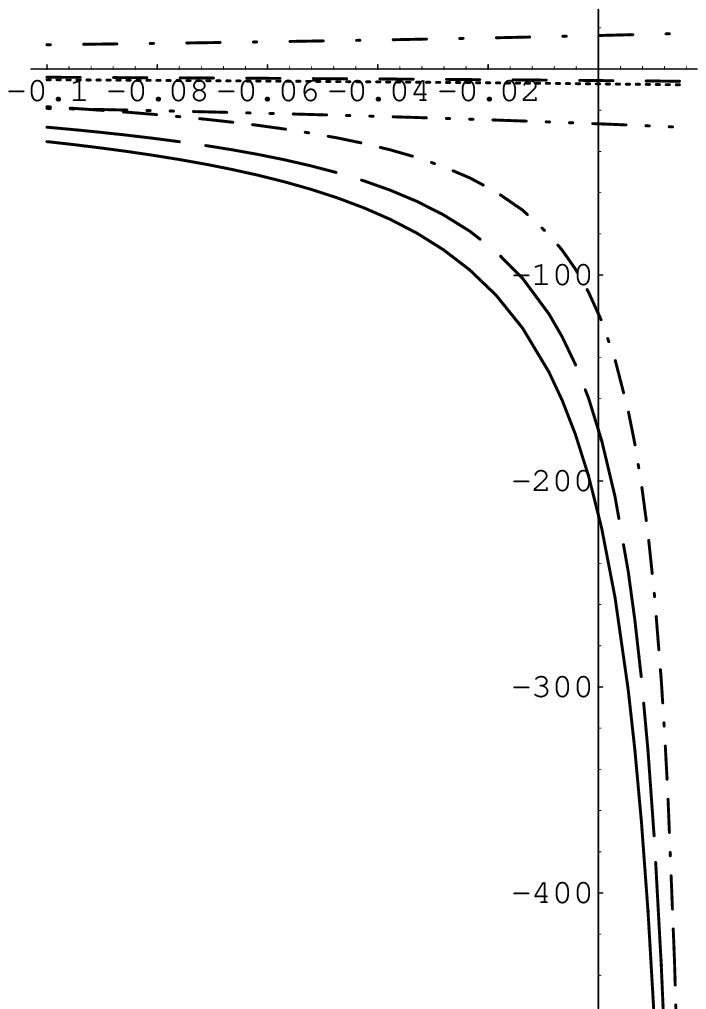,width=7.0cm}}}
\put(200,0){\makebox(100,120){\epsfig{file=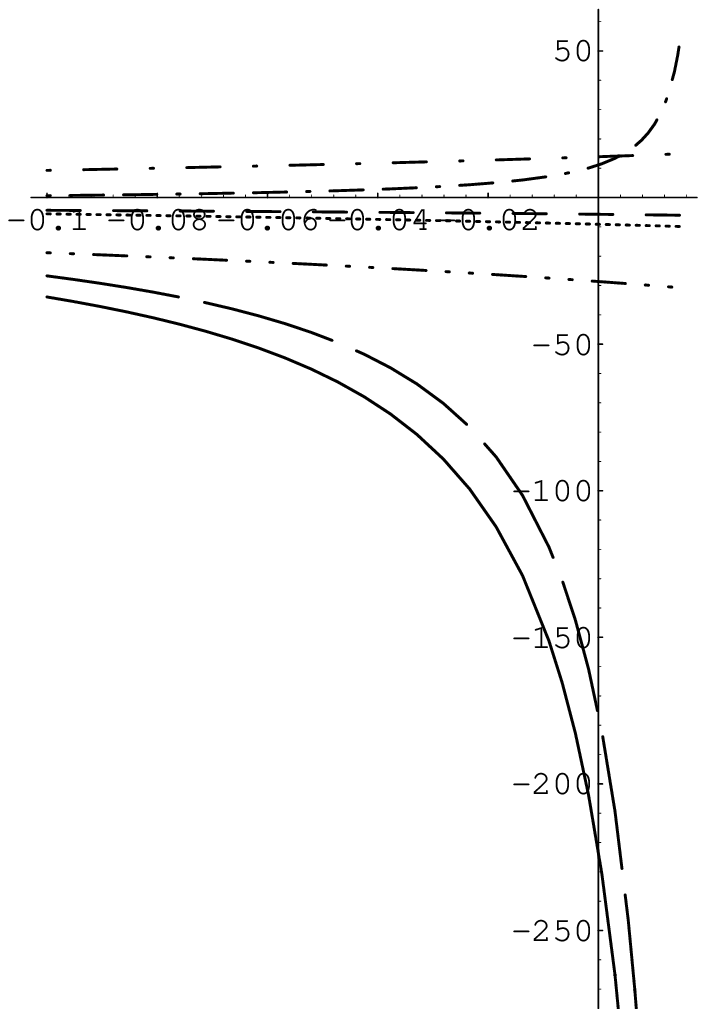,width=7.0cm}}}
\put(50,200){$a)$}
\put(250,200){$b)$}
\put(50,-60){$c)$}
\put(250,-60){$d)$}
\put(117,400){\scriptsize{$q^2 \, [GeV^2]$}}
\put(317,400){\scriptsize{$q^2 \, [GeV^2]$}}
\put(117,140){\scriptsize{$q^2 \, [GeV^2]$}}
\put(317,115){\scriptsize{$q^2 \, [GeV^2]$}}
\end{picture}
\vskip 2.8cm

Figure 8

\end{figure}

\end{center}

\end{document}